\documentclass[twocolumn,twocolappendix]{aastex631}
\usepackage{xcolor}
\usepackage{graphicx}
\usepackage{listings}
\usepackage{amsmath}
\usepackage{comment}

\definecolor{mauve}{rgb}{0.88, 0.69, 1.0}
\definecolor{oldmauve}{rgb}{0.4, 0.19, 0.28}

\newcommand{\insitu}{{\it in situ}}
\newcommand{\MH}{\ensuremath{\mathrm{[M/H]}}}
\newcommand{\Teff}{\ensuremath{\mathrm{T}_{\rm eff}}}
\newcommand{\aFe}{\ensuremath{\mathrm{[\alpha/M]}}}
\newcommand{\GBP}{\ensuremath{G_\mathrm{BP}}}

\shorttitle{The Poor Old Heart of the Milky Way}
\shortauthors{Rix et al.}

\begin{document}

\title{The Poor Old Heart of the Milky Way}

\author[0000-0003-4996-9069]{Hans-Walter Rix}
\affiliation{Max-Planck-Institut f{\"u}r Astronomie, K{\"o}nigstuhl 17, D-69117 Heidelberg, Germany}

\author[0000-0002-0572-8012]{Vedant Chandra}
\affiliation{Center for Astrophysics $\mid$ Harvard \& Smithsonian, 60 Garden St, Cambridge, MA 02138, USA}

\author[0000-0001-8006-6365]{Ren\'e Andrae}
\affiliation{Max-Planck-Institut f{\"u}r Astronomie, K{\"o}nigstuhl 17, D-69117 Heidelberg, Germany}

\author[0000-0003-0872-7098]{Adrian M. Price-Whelan}
\affiliation{Center for Computational Astrophysics, Flatiron Institute, 162 Fifth Avenue, New York, NY 10010, USA}

\author[0000-0002-0572-8012]{David H. Weinberg}
\affiliation{Department of Astronomy and Center for Cosmology and AstroParticle Physics, The Ohio State University, Columbus, OH 43210, USA}
% \affiliation{Institute for Advanced Study, Princeton, NJ 08540, USA}

\author[0000-0002-1590-8551]{Charlie Conroy}
\affiliation{Center for Astrophysics $\mid$ Harvard \& Smithsonian, 60 Garden St, Cambridge, MA 02138, USA}

\author[0000-0003-4996-9069]{Morgan Fouesneau}
\affiliation{Max-Planck-Institut f{\"u}r Astronomie, K{\"o}nigstuhl 17, D-69117 Heidelberg, Germany}

\author[0000-0003-1879-0488]{David W Hogg}
\affiliation{Center for Cosmology and Particle Physics, Department of Physics, New York University, 726
Broadway, New York, NY 10003, USA}
\affiliation{Center for Computational Astrophysics, Flatiron Institute, 162 Fifth Avenue, New York, NY 10010, USA}
\affiliation{Max-Planck-Institut f{\"u}r Astronomie, K{\"o}nigstuhl 17, D-69117 Heidelberg, Germany}

\author[0000-0003-1879-0488]{Francesca De Angeli}
\affiliation{Institute of Astronomy, University of Cambridge, Madingley Road, Cambridge, CB3 0HA, UK}

\author[0000-0002-1590-8551]{Rohan P. Naidu}
\altaffiliation{NASA Hubble Fellow}
\affiliation{MIT Kavli Institute for Astrophysics and Space Research, 77 Massachusetts Ave., Cambridge, MA 02139, USA}

\author[0000-0002-5818-8769]{Maosheng Xiang}
\affiliation{National Astronomical Observatories of China, Chinese Academy of Sciences, Beijing, 100012, China}

\author[0000-0002-9455-157X]{Daniela Ruz-Mieres}
\affiliation{Institute of Astronomy, University of Cambridge, Madingley Road, Cambridge, CB3 0HA, UK}

%\author{contributing friends and colleagues}

\correspondingauthor{Hans-Walter Rix}
\email{rix@mpia.de}

\begin{abstract}
Massive disk galaxies like our Milky Way should host an ancient, metal-poor, and centrally concentrated stellar population. This population reflects the star formation and enrichment in the few most massive progenitor components that coalesced at high redshift to form the \emph{proto-Galaxy}. While metal-poor stars are known to reside in the inner few kiloparsecs of our Galaxy, current data do not yet provide a comprehensive picture of such a metal-poor ``heart'' of the Milky Way. We use information from Gaia DR3, especially the XP spectra, to construct a sample of 2 million bright ($\GBP<15.5$\,mag) giant stars within $30^\circ$ of the Galactic Center with robust \MH\ estimates, $\delta\MH\lesssim 0.1$. For most sample members we can calculate orbits based on Gaia RVS velocities and astrometry. This sample reveals an extensive, ancient, and metal-poor population that includes $\sim 18,000$ stars with $-2.7<\MH<-1.5$, representing a stellar mass of $\gtrsim 5\times 10^7$\,M$_\odot$. The spatial distribution of these $\MH<-1.5$ stars has a Gaussian extent of only $\sigma_{\mathrm{R_{GC}}} \sim 2.7$~kpc around the Galactic center, with most of these orbits being confined to the inner Galaxy.  At high orbital eccentricities, there is clear evidence for \emph{accreted} halo stars in their pericentral orbit phase.
%Massive disk galaxies like our Milky Way should have an ancient, metal-poor ($\MH <-1.5$), and centrally concentrated \emph{in-situ} spheroidal stellar population, reflecting the first star formation and enrichment in the earliest, most massive Milky Way progenitor component. 
%The metallicity distribution rises steeply from $\MH\lesssim -2.5$ to $-2$, and then has a slope of $\mathrm{d\,log\,N[M/H]}\,/\,\mathrm{d\,[M/H]}\approx -1$ from $\mathrm{[M/H]} < -2$ to $-1$. 
Stars with $\MH< -2$ show no net rotation, whereas those with $\MH\sim -1$ are rotation dominated.
%, as expected for self-enrichment with an ample and constant gas supply. 
Most of the tightly bound stars show $[\alpha/\text{Fe}]$-enhancement and [Al/Fe]--[Mn/Fe] abundance patterns expected for an origin in the more massive portions of the proto-Galaxy. 
  %The level of rotational support in this population steadily increases with \MH, with a net prograde motion for all $\mathrm{[M/H]}>-2$. 
These central, metal-poor stars most likely predate the oldest part of the disk ($\tau_{\text{age}}\approx 12.5$~Gyrs), which implies that they formed at $z\gtrsim 5$, forging the proto-Milky Way.
\end{abstract} 


\section{Introduction} \label{sec:intro}

%\noindent\textcolor{orange}{{\bf Open Issues/Remaining To-Do's}}
%\begin{itemize}
%    \item Do we simply publish the sample, or refer to Andrae+ 2022?
%    \item Revise Abstract
%   \item Qantify the spatial extent of poor, heart population
%    \item How clear is the argument that Kraken/Koala/Hercules are not distinct events but tails of the in-situ distribution?
%    \item Have we properly acknowledged what was known before? [Abstract?]
%\end{itemize}

Understanding the formation history of our own Galaxy, especially its earliest phases, has been a central goal of \emph{Galactic Archeology} for decades  \citep{Freeman_BlandHawthorne_2002}, with dramatic progress enabled by a suite of spectroscopic surveys \citep[e.g.,][]{Yanny2009,DeSilva2015,Majewski2017,Conroy2019b} and ESA's Gaia mission \citep{Perryman2001,GaiaDR3}. The earliest phases of the  Milky Way's star-formation and enrichment history are reflected at the present epoch in the orbit- and abundance- distribution of old and metal-poor stars. In the context of the hierarchical formation of massive disk galaxies like the Milky Way, we should expect the oldest and most metal-poor stars (say $\MH < -1.5$) to be a mix of stars that a) formed within one of the main overdensities which coalesced early to form the proto-Galaxy, or b) formed early in distinct satellite galaxies that eventually merged with the main body \citep{Zolotov2009}. The first channel
is commonly referred to as \insitu~ formation, the second \emph{accreted} \citep{Tumlinson2010, Pillepich2015, El-Badry2018a,Renaud2021b}. If the \insitu~ stars formed in a considerably more massive potential well than the \emph{accreted} stars, this difference in origin should also be reflected in their stars' abundance patterns, such as [$\alpha$/Fe] vs. [Fe.H] or [Mg/Mn] vs. [Al/Fe] \citep[][and others]{Zolotov2010,Hawkins2015,Das2020,Horta2021,Belokurov2022,Conroy2022}.

However, at very early epochs, this distinction between \insitu\ and \emph{accreted} may become rather blurry. High-resolution formation simulations of galaxies like the Milky Way show \citep[e.g.][]{Renaud2021a,Renaud2021b} that pieces of comparable mass may rapidly coalesce early on (say, $z>4$) in a sequence of ``major mergers''. One terminology choice is to label only the one piece that was somewhat more massive as the \insitu~ component,  and to label all other pieces as accreted (which put together may constitute the majority of mass). However, simulations show that it may be operationally impossible later on to differentiate what was \insitu\ and what accreted at very early epochs, whether one uses orbit or abundance information \citep[e.g.][]{Renaud2021b,Brauer2022,Orkney2022}. Alternatively, one could label collectively all the major pieces that coalesced very early on, say at $z\gtrsim 5$, as the \emph{proto-Galaxy}. \citep[see][]{Conroy2022}\footnote{\cite{Belokurov2022} use the term ``young Galaxy''}, and then differentiate subsequent additions at slightly later epochs (say, $z<3$) as either \insitu, if the material was brought in as gas, or as \emph{accreted}, if the stars formed in a distinct potential well that then subsequently merged. Here, we opt for the second terminology, in part in light of the results we find. Consequently, we refer to the oldest parts of the Milky Way 
%that may have merged without ever orbiting as distinct entities 
as the \emph{proto-Galaxy}, without trying to single out one of the contributing pieces as \insitu .

%Simulations do not provide definitive predictions about the fraction of all old, low-\MH\ stars that are \insitu (vs. accreted), and this fraction is expected to vary strongly with both the age and \MH\ of the stars. There are also no definitive predictions about the spatial distribution of old, metal-poor \insitu stars: most old and low-\MH\ \insitu stars originally formed quite deep inside the main potential well but may have become dynamically heated in the subsequent evolution \citep[e.g.,][]{Buck2018,Buck2019,ElBadry2018_oldest_stars}.  There is broad agreement among simulations that the fraction of accreted old, low-\MH\ stars increases with orbital radius or binding energy \citep[e.g.][]{Tumlinson2010, Pillepich2015}.

%The paragraph below felt out of place - could be moved to later in the Intro, or later in the paper.  Might not be needed at all.
%For the old, metal-poor and kinematically hot population of stars at $R_{\text{GC}}\gtrsim R_\odot\sim$8~kpc the term \emph{stellar halo} is well established. In contrast, there is no terminological consensus on whether such old, metal-poor and centrally concentrated \insitu stars in the Milky Way or other galaxies should be denoted as in the ``inner halo'' or the metal-poor part of the bulge; here, we shall use the relatively agnostic term ``metal-poor spheroid'' \citep[e.g.,][]{Howes2015,Ness2016,Barbuy2018,Arentsen2020a}.

The nature of the accreted component of our Galaxy at $R_{\text{GC}}\gtrsim 5$~kpc has come into focus due to the combination of Gaia and large ground-based spectroscopic surveys. Tidal debris from the disrupted Sagittarius satellite \citep{Ibata1994} dominates the halo from $20-50$~kpc with stars mostly on highly inclined orbits with substantial angular momentum \citep{Majewski2003, Naidu2020}; at $5-25$~kpc the debris from the disrupted Gaia-Sausage-Enceladus (GSE) satellite \citep{Helmi2018,Belokurov2018a,Naidu2020} dominates, with stars on highly eccentric orbits ($e\gtrsim 0.7$) with little (slightly retrograde) angular momentum. Beyond these dominant components, a growing number of additional, distinct accreted components of the Galaxy have been identified \citep[e.g.,][]{Newberg2009,Kruijssen2019,Myeong2019,Naidu2020,Yuan2020,Malhan2022a}.  Beyond their orbits, the abundance patterns of these stars also point toward an accreted origin: their distribution in the 
\aFe-\MH\  plane (lower \aFe\  at a given \MH) indicates that their birth material was enriched in a potential well of lower mass, i.e. a satellite galaxy \citep[e.g.][]{Frebel_ARAA,Hawkins2015,Lee2015}.  In addition to presumed satellite galaxy debris, the stellar halo also appears to encompass a large number of disrupted star clusters \citep[e.g.][]{Malhan2018a,Shipp2018,Bonaca2021}.

There has also been progress in understanding the old proto-Galactic or  \insitu\ component of the Milky Way. The old and metal-poor Galactic \insitu\ component that has been mapped best is the old, $\alpha$-enhanced disk \citep[e.g.,][]{Hayden2015, Bonaca2020, Belokurov2020, XiangRix2022}: its stars seem to date back to $\ge 12.5$~Gyrs (or $z\gtrsim 5$), are centrally concentrated and form a thick disk. However, the old $\alpha$-enhanced disk metallicity distribution function (MDF) is truncated, or at least drops sharply, below $\MH=-1$. It seems inevitable that there must have been a substantive proto-Galactic stellar population responsible for enriching the old disk's birth material to $\MH=-1$. Recent surveys have provided clear evidence for a so-called \insitu~halo component in the Milky Way \citep[e.g.,][]{Bonaca2017,Haywood2018, DiMatteo2019,Naidu2020,Bonaca2020,Belokurov2020,Horta2021}. Many of these studies drew on samples at the Solar radius and beyond, and those \insitu~ halo stars may reflect very early disk stars that were kicked up, or splashed to hotter orbits.

While the chemodynamical structure of the Galaxy at $R_{\text{GC}}\gtrsim 5$~kpc has come into focus over the past decade, the nature of the inner Galaxy -- especially at low metallicity -- has proven more elusive. Fortunately, recent work has begun to piece together a picture of the metal-poor inner Galaxy with limited sample sizes \citep[e.g.,][]{Ness2013,GarciaPerez2013,Schlaufman2014,Casey2015,Ness2015,Koch2016,GarciaPerez2018,Reggiani2020,Arentsen2020a,Lucey2021}. In particular, narrow-band photometric surveys have spectacularly enabled a more efficient selection of metal-poor stars towards the inner Galaxy \citep[e.g.,][]{Arentsen2020a,Arentsen2020b}, providing metal-poor samples in the inner Galaxy of several thousand objects. Spectroscopic follow-up has shown that these stars are indeed metal poor and show very little rotation at the lowest metallicities \citep{Arentsen2020b}. \citet{Kruijssen2019,Kruijssen2020} and \citet{Forbes2020} recently used the properties of inner Galaxy globular clusters as tracers of a very old Milky Way component, which they dubbed \emph{Kraken} and \emph{Koala}. They considered these possible instances of early accretion.

Most recently, \citet{Belokurov2022} and \citet{Conroy2022} have been able to shed first light on the question of how the old disk emerged from an earlier, more metal-poor population of stars on dynamically hot orbits.  \citet{Belokurov2022} combined APOGEE and Gaia astrometry to study the kinematics of stars in the inner Galaxy down to metallicities of $\MH \sim -1.5$. They found that metal-poor stars in the inner Galaxy with abundance patterns they identified with \insitu~ formation (dubbed \emph{Aurora}) show rather little net rotation, and reasoned that these stars preceded the Milky Way's ancient disk \citep[see also][]{Arentsen2020a}.  \citet{Conroy2022} used chemistry, kinematics, and ages from H3 and Gaia data to identify likely proto-Galactic stars down to $\MH\sim -2.5$ with ages $\gtrsim13$ Gyr. This recent work clearly implies that our Milky Way has some form of a proto-Galactic population. Whether or not there exists an operationally separable accreted population buried deep in the potential well of our Galaxy remains an open question \citep[cf.][]{Kruijssen2020, Horta2021, Myeong2022}.

It is on this background that we aim to flesh out a more comprehensive picture of the abundance-orbit distribution of metal-poor stars in the inner Galaxy, $R_{\rm{GC}}\lesssim 5$~\rm{kpc}. Here, sensible definitions of metal-poor could be a) more metal-poor than the oldest disk stars, i.e. $\MH < -1$, or b) more metal poor than large published samples of stars with abundances and orbits in the inner Galaxy, i.e. $\MH < -1.5$.

Specifically, we carry out a comprehensive search for metal-poor stars towards the Galactic center, drawing on the immense wealth of new information that Gaia DR3 now affords. Our work relies mainly on newly published low-resolution BP/RP spectra \citep[hereafter collectively referred to  as `XP', ][]{2021A&A...652A..86C,DeAngeli2022} and photometry synthesized from these spectra \citep{Montegriffo2022b}. In principle, various metallicity estimates exist for millions of stars in DR3. Yet, the XP-based estimates from \citet{Andrae_XP_DR3} suffer from strong biases,\footnote{The metallicity biases reported in \citet{Andrae_XP_DR3} presumably originate from systematics in the \emph{ab initio} SED models and from our imperfect understanding of the XP instrument used to transfer those model SEDs into XP spectra.} while the RVS-based metallicity estimates from \citet{Recio-Blanco-RVS-abundances} are currently limited to bright stars with high-quality Gaia RVS spectra. 

%As Gaia's initial and published \MH\ values have been based exclusively on \emph{ab initio} models, they contain a significant -- in some regimes large -- a fraction of \MH\ estimates that are discrepant with high-quality spectral estimates. This is a potentially prohibitive shortcoming for rare object searches (metal-poor stars in the inner Galaxy).
For our analysis, we select giant stars within 30$^\circ$ of the Galactic center, derive data-driven \MH\ estimates from their XP spectra, and combine them with RVS velocities to obtain orbits. This provides a sample of 1.5~million stars in the inner Galaxy, for a chemodynamical study.

The rest of the paper is organized as follows: in Section~\ref{sec:data} we describe the initial selection of the sample, and the subsequent data-driven derivation of \MH\ estimates, using their XP spectra and AllWISE photometry \citep{2014yCat.2328....0C}. In Section~\ref{sec:results} we present the spatial, \MH~ and orbit distribution of these stars, 
along with the abundance patterns for a small subset. In Section~\ref{sec:discussion} we summarize these results, put them in the context of Galactic archaeology, and sketch a few avenues for follow-up work.

\section{Data Sets and Derived Abundance and Orbit Quantities}\label{sec:data}

We aim to devise a well-defined and large sample of stars in the ``inner Galaxy'' that consists of stellar tracers that cover the full age and [M/H] range with comparable selection effects.
For that we need tracers that are luminous enough to reach the distance of the Galactic center and beyond, even in the presence of some dust extinction. 
For \emph{chemo-dynamical} mapping, we need a robust and precise estimate of \MH\ and an estimate of the 6D phase space coordinates ($\mathit{x,v})$ to estimate the orbits. Ideally, we would also like to have precise age estimates and individual abundances [X/H], or at least an estimate of the $\alpha$-enhancement, $[\alpha /M]$. Using age estimates and individual abundances for these giant stars is beyond the scope of this work, as is a full accounting of the sample selection function.

To build such a sample, we first identify all likely red giant stars  (RGB and RC) in the direction of the Galactic center that: a) are bright enough that we can expect a robust and precise estimate of \MH\ from Gaia XP data \citep{DeAngeli2022,Montegriffo2022a}; and b) have radial velocities from Gaia RVS \citep{Katz_RVS_DR3} to estimate the stars' orbits in conjunction with their parallaxes, $\varpi$, and proper motions, $\vec{\mu}$. Then we re-derive \MH~ estimates for these stars from their XP spectra with a data-driven approach that draws on the SDSS's APOGEE survey \citep[DR17,][]{SDSS_DR17}, to address the documented shortcomings of the {\tt GSP-Phot} abundance estimates in Gaia DR3 \citep{Andrae_XP_DR3}.
Finally, we calculate the orbits of the sample members from their ($\mathit{x,v}$), which requires a model for the Galactic potential, $\Phi(\mathit{x})$ \citep[here][]{gala}.

\subsection{Initial Gaia Query}

The initial \emph{inner Galaxy sample} was devised via the following ADQL query:

\begin{lstlisting}
select  * from gaiadr3.gaia_source 
where
parallax < 100.*power(10.,0.2*(0.9 - (phot_g_mean_mag - 1.5*(bp_rp-1.))))
and parallax < 1.
and abs(b)<30 and (l<30 or l > 330.)
and bp_rp between 1.0 and 3.5
and phot_bp_mean_mag < 15.5
and has_xp_continuous='true'
\end{lstlisting}

The first condition on the parallax selects stars whose absolute magnitude is more luminous than $M_G=0.9$, aimed at selecting giant stars at luminosities that just include the red clump (RC); the {\tt -1.5*(bp$\_$rp-1.)} term reflects an approximately dereddened magnitude. The query is articulated in such a way that zero or negative parallaxes are accommodated. The second parallax condition only excludes the immediate ``foreground'' at $D<1$~kpc. The conditions on $l$ and $b$ select a rectangular region of 30$^\circ$ around the Galactic center. The lower limit on bp\_rp eliminates most luminous hot stars for which \MH -estimates are not feasible (but not all in the presence of reddening). The condition on phot\_bp\_mean\_mag is designed to select objects bright enough in the blue (BP) part of the XP spectra for robust \MH -estimates; it still allows to select RC stars at the distance of the Galactic center in the presence of moderate reddening. The final line in the selection ensures that continuous XP spectra are available. This query yields 2.1 million objects.

\subsection{Deriving Robust \MH\ Estimates}

% \begin{figure}[h!]
% \begin{center}
% \includegraphics[width=\columnwidth]{dust_validation.pdf}
% \end{center}
% \caption{(In-)sensitivity of [M/H] estimates to extinction, $A_K$. The Y-axis shows the difference between \MH ~estimates based on \emph{xgboost} and on APOGEE as a function of APOGEE's $A_K$. There is no evidence for any systematic trend of $\Delta\MH$ with $A_K$, to an extinction level that corresponds to $A_V \approx 3$.}
% \label{fig:dust_validation}
% \end{figure}

\begin{figure}%[h!]
\begin{center}
\includegraphics[width=\columnwidth]{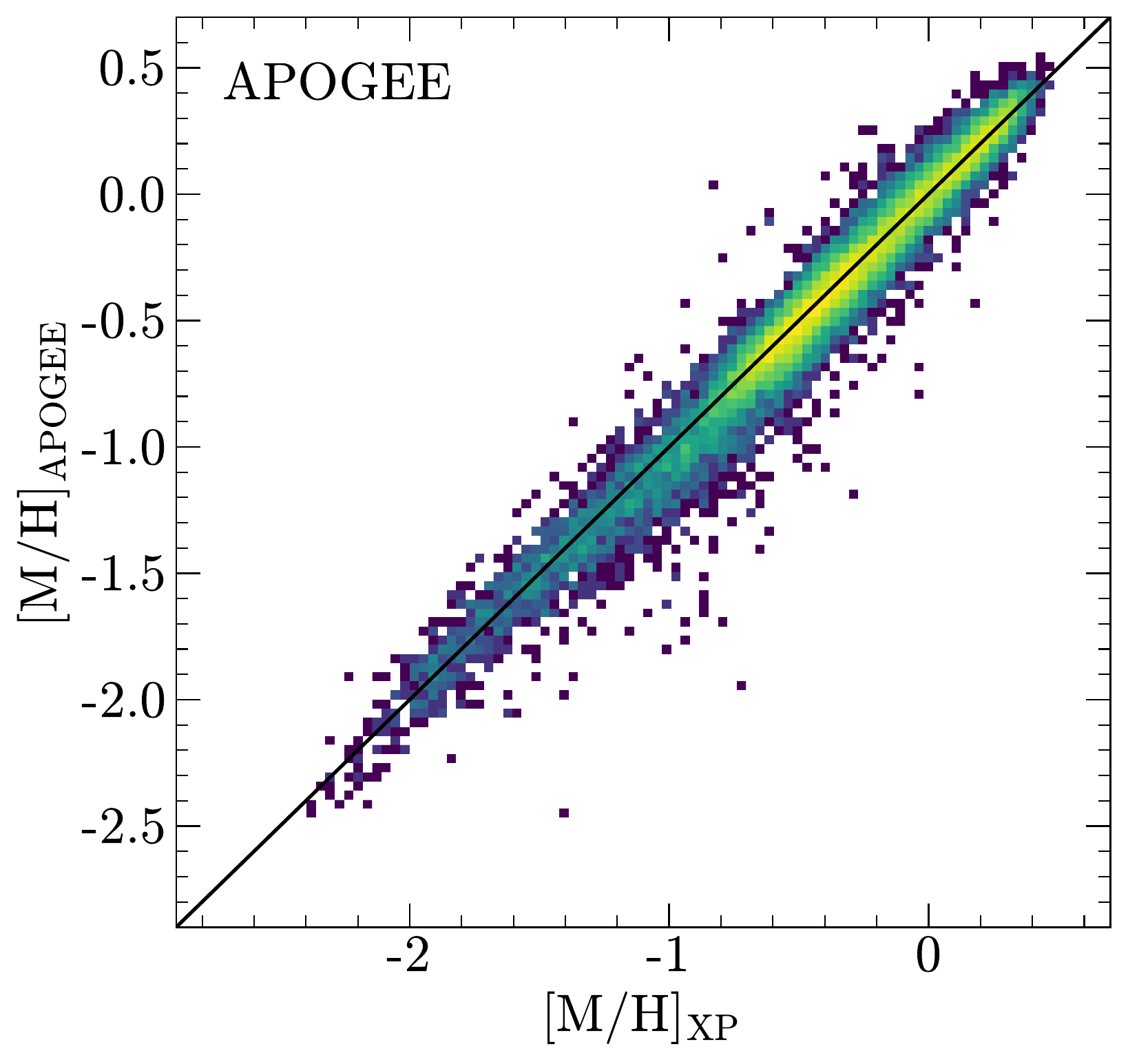}
\includegraphics[width=\columnwidth]{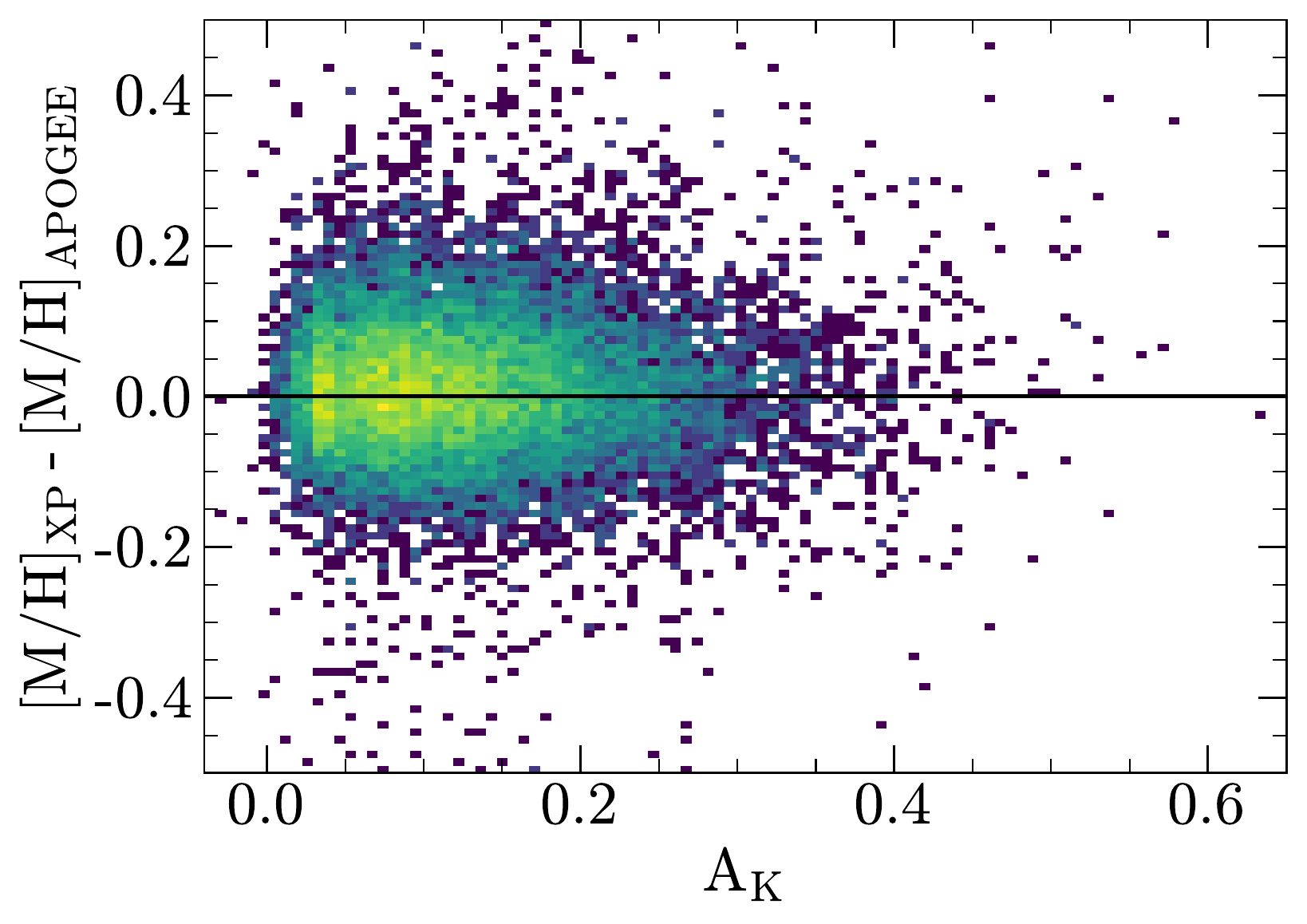}
\end{center}
\caption{Top: Validation of our \MH\ estimates from \textit{Gaia} XP data as function of actual [M/H] from APOGEE.
Bottom: (In-)sensitivity of \MH\ estimates to extinction, $A_K$. The Y-axis shows the difference between \MH\ estimates based on XP spectra and APOGEE as a function of APOGEE's $A_K$. There is no evidence for any systematic trend of $\Delta\MH$ with $A_K$ to an extinction level that corresponds to $A_V \approx 3$.}
\label{fig:MH_validation}
\end{figure}

% \begin{figure*}%[t!]
% \centering

% \caption{}
% \label{fig:cross-validation}
% \end{figure*}

 While about 75\% of this sample has RVS velocities in Gaia DR3, only a small fraction of them have metallicities or abundances derived from the RVS spectra \citep{Recio-Blanco-RVS-abundances}. This leaves the route of estimating \MH\ from the low-resolution XP spectra of these objects. Consequently, our objective is to train a machine-learning algorithm that can precisely, accurately, and robustly predict \MH\ from Gaia DR3 data.

 Estimating \MH\ based on machine learning involves four aspects. 1) It involves the choice of the training sample, for which we adopt SDSS DR17 \citep[APOGEE,][]{2022ApJS..259...35A} since its abundances are well validated, it covers the inner Galaxy and contains mainly giants, many with high extinctions. 2) It involves the choice of the data features on which to train the prediction; these can be derived from the XP spectra but may also entail, as we discuss below, external data available across the sample, such as near-infrared photometry from ALLWISE \citep{2014yCat.2328....0C}. 3) It involves the choice of the machine learning algorithm, for which we adopt the extreme gradient boosting algorithm \citep[][hereafter XGBoost]{Chen:2016:XST:2939672.2939785}. XGBoost is one of the best algorithms currently available; it is straightforward and computationally inexpensive to train and it can outperform other methods such as deep learning \citep[e.g.][]{https://doi.org/10.48550/arxiv.2207.08815}. 4) and finally, we need to validate our \MH\ prediction, which then speaks to the quality of the data, the training sample, and the algorithm.

%\begin{figure}[h!]
%\begin{center}
%\includegraphics[width=8cm]{aM_validation.pdf}
%\end{center}
%\caption{Validation of [$alpha$/H]. \textcolor{orange}{Eliminate? It would need VERY careful cross-validation.} \textcolor{red}{Since we're not using alpha in our arguments here, I'd likewise vote to eliminate this. As Rene has shown, a lot of this could be from correlations.}}
%\label{fig:aM_validation}
%\end{figure}

The choice of the correct data features to train and predict \MH from the XP spectra is not trivial. One obvious option is to directly use the $2\times 55$ modified Gauss-Hermite coefficients that describe XP spectra. But our numerical experimentation implies that other or additional data features may yield more precise and robust results with a machine learning approach\footnote{If stellar features are to be estimated from XP spectra in a full forward-modeling approach, the situation may be different.}, for several pragmatic reasons. Much of the information about narrow metal line features is contained in high-order coefficients, which are often noisy. We have prior information on which parts of the spectrum are highly diagnostic in \MH\ estimates and which ones are not; we need not ask an algorithm to learn this. Any estimate of \MH\ requires an implicit estimate of the effective temperature \Teff, which is covariant with \MH; and \Teff\ is also highly covariant with reddening. Therefore, providing additional information that helps break that degeneracy is precious. The near-infrared photometry from ALLWISE \citep{2014yCat.2328....0C} exists across the sky and proves powerful in this respect. 

In practice, we follow the approach of \citet{Montegriffo2022b} to calculate synthetic photometry from the XP spectra in a wide range of mostly narrowband filters, in particular those filters with established utility in identifying metal-poor stars: Str\"omgren filters \citep{Stromgren_Filters}, JPAS+ filters \citep{JPAS} and Pristine {\it H\,}\&{\it \,K} filters \citep{Pristine}. On this basis, we use XGBoost to train the prediction of \MH\ using the \emph{entire} SDSS DR17 APOGEE of giants with $G_\textrm{BP}<15.5$ ($\sim 230,000$ stars). Details are given in the Appendix. 

Training on a particular data set and using a model that is discriminative, rather than generative, raises several issues.  First, we (obviously) tie our results to the metallicity scale of SDSS DR17. Second, estimates of \MH~ may be  
drawn into the support of the training set. In particular, XGBoost will not extrapolate \MH\ significantly beyond the APOGEE DR17 range, as it is a tree-based method that segments the feature space and assigns a mean label to each such segment. Finally, generative data-driven spectral models, such as \emph{The Cannon} \citep{Ness2015}, will presumably degrade more gracefully towards low signal to noise.  Here we have addressed the last point implicitly by restricting the sample to $G_{\text{BP}}<15.5$.

This results in \MH\ estimates for about 2 million giants toward the inner Galaxy, 1.58 million of which have RVS velocities, and 1.25 million have both RVS velocities and $\varpi/\sigma_\varpi \ge 5$, which is our minimal condition for an orbit estimate.
For subsequent analysis, we have eliminated 2\% of the sample as reddened hot stars, which can be readily recognized by their position in the $m_1$--{\tt bp$\_$rp} color plane, 
where $m_1\equiv v-2b+y$ is the metallicity-sensitive Str\"omgren filter combination \citep[see][for  details]{Stromgren_Filters}; specifically, we eliminate sources with $m_1 + 0.16\, (\mathtt{bp\_rp}-1) < 0$ and $\mathtt{bp\_rp} < 2.7$.

\subsection{\MH\ Validation}\label{sec:MHvalidation}

We can explore and validate the precision and accuracy of these \MH\ estimates through various comparisons with analogous estimates from (higher-resolution) spectroscopic surveys. For this validation, we consider SDSS~APOGEE \citep{SDSS_DR17}, LAMOST \citep[here][]{Xiang2019}, GALAH \citep[DR3][]{GALAH_DR3}, and Gaia's {\tt GSP-Spec} \citep{Recio-Blanco-RVS-abundances}.

We start by considering the $\sim 17,000$ stars in common between APOGEE and our sample, shown in Figure \ref{fig:MH_validation} (top panel). 
This panel of Figure \ref{fig:MH_validation} illustrates that for this sample, XGBoost provides a remarkably precise ($\lesssim 0.1$ in the median), accurate and robust \MH ~estimate across $-2.5<\MH <0.5$. It should be noted that XGBoost was trained on the \emph{full} all-sky APOGEE sample. So this is a validation on a small subset of the full training set, not a pure cross-validation. The accuracy of the \MH~ prediction (i.e. lack of systematic offset) is therefore by construction. It is particularly remarkable that basically all stars predicted to be metal-poor from $\MH _{\text{XP}}$ are indeed metal-poor according to \MH$_{\text{APOGEE}}$: a selection of, e.g., metal-poor objects by $\MH_{\text{XP}}$ appears to be nearly pure.  The bottom panel of Figure \ref{fig:MH_validation} shows that these \MH\ estimates remain unbiased, at a level of a few percent, even in the presence of substantial dust extinction $A_K=0.3$, corresponding to about 3~magnitudes of $A_V$.

Appendix~\ref{sec:more_mhvalidation} details a true cross-validation of our $\MH_{\text{XP}}$ estimates against three external data sets: LAMOST, GALAH and GSP-Spec. Broadly speaking, the cross-validation affirms that the \MH$_{\text{XP}}$ estimates are precise and robust (enabling pure \MH-based sample selection). Our estimates have a root mean square difference $\approx 0.1$~dex compared to these datasets, with minimal bias. Most importantly, the performance remains unbiased toward the metal-poor end, affirming our confidence in selecting metal-poor stars based on XP metallicities. We refer to Appendix~\ref{sec:more_mhvalidation} for more detailed validation of the [M/H] estimates. 

\subsection{Orbits of the Sample Members}\label{sec:orbit_calculation}

For sample members with suitable 6D phase-space information, we calculate their orbits. The sky positions and proper motions are available for all of them, radial velocities, and good parallax measurements only for a good fraction of the sample. In practice, we take all stars that have RVS velocities with $\delta v_\textrm{RVS}<5~\textrm{km}~\textrm{s}^{-1}$ and $\varpi/\sigma_\varpi > 5$, about 1.25 million objects across all \MH. 

To compute orbits and compute orbital properties, we use a four-component Milky Way mass model consisting of spherical Hernquist nucleus and bulge components, an (approximate) exponential disk component \citep{Smith:2015}, and a spherical NFW halo component.
We adopt a radial scale length $h_R=2.6~\textrm{kpc}$ and scale height $h_z=300~\textrm{pc}$ \citep{Bland-Hawthorn:2016}, and fit for the masses and scale radii of the nucleus, bulge, and halo components using the same compilation of Milky Way enclosed mass measurements as used to define the \texttt{MilkyWayPotential} in \texttt{gala} \citep{gala, adrian_price_whelan_2020_4159870}, with the additional constraint of having a circular velocity at the solar position $v_c(R_0) = 229~\textrm{km}~\textrm{s}^{-1}$ \citep{Eilers:2019}.
We compute actions using the ``St\"ackel Fudge'' \citep{Binney:2012, Sanders:2012} as implemented in \texttt{galpy} \citep{galpy}.
To compute the orbital eccentricity, pericenter, and apocenter values, we numerically integrate the orbits with a timestep of 0.5~Myr across four times the radial orbit period (estimated by the computed orbital frequencies from the action solver).

The \MH~ estimates for 1.5 million stars towards the Galactic center, along with orbits for the RVS subsample of 1.25 million stars, are available as supplementary material to this article and are hosted online\footnote{\url{https://doi.org/10.5281/zenodo.7035809}} \citep{hans_walter_rix_2022_7035810}. 

\section{Results}\label{sec:results}

We now present the properties of the low-\MH\ stellar population contained within this sample. We start with the spatial distribution, then move to the \MH-distribution, the orbit distribution, and finally to a first exploration of the abundance patterns of the metal-poor stars, especially the level of $\alpha$-enhancement. This is all done with an eye toward what we can learn about a central metal-poor \insitu\ halo or bulge population.
%DW Maybe "halo/bulge population".  I removed the commas because that punctuation implies that bulge is a different term for halo, which I don't think is what you intend here.

\subsection{Spatial Distribution of Metal-Poor Stars}

\begin{figure*}[!t]
\centering
\includegraphics[width=5.9cm]{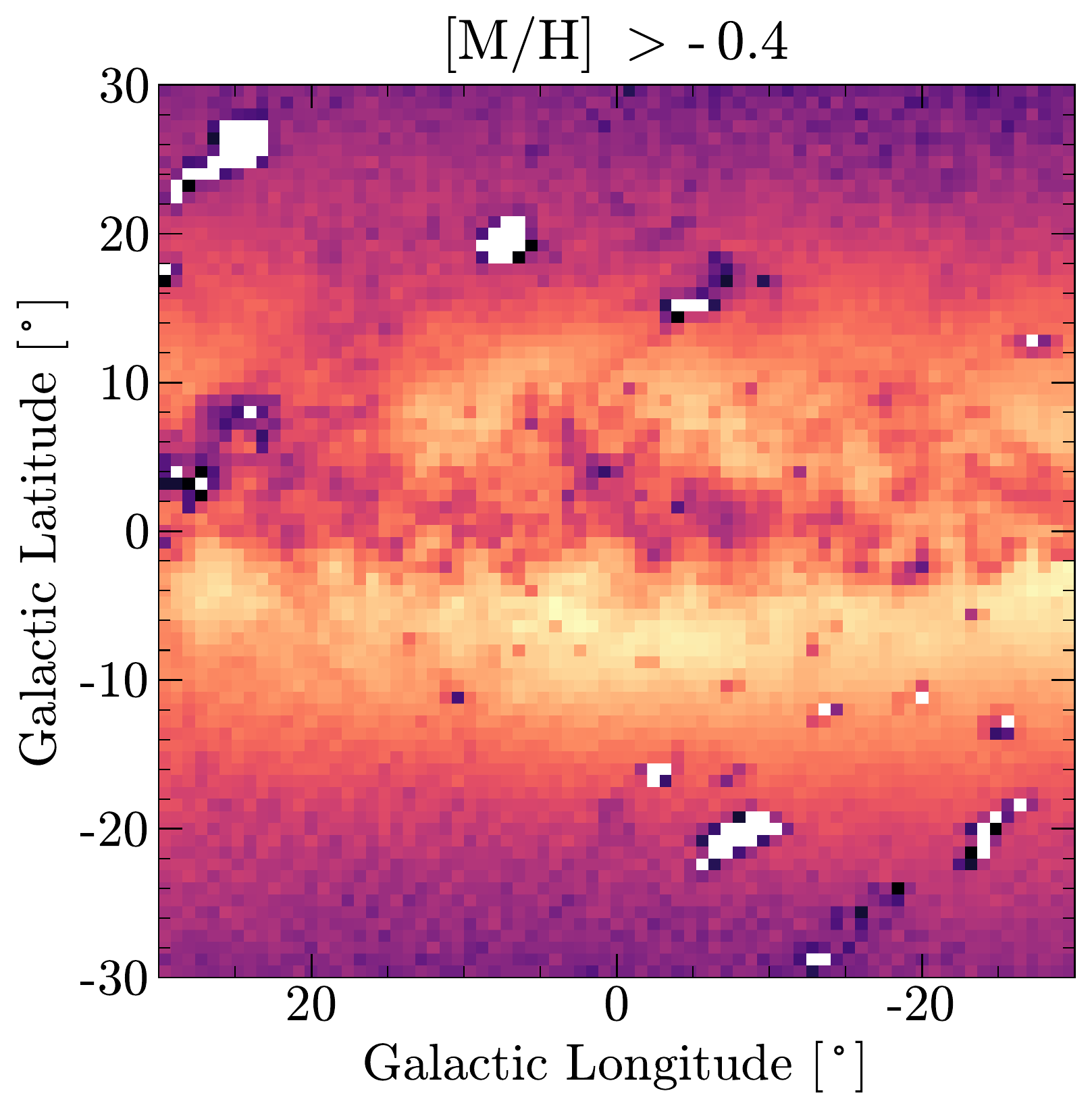}
\includegraphics[width=5.9cm]{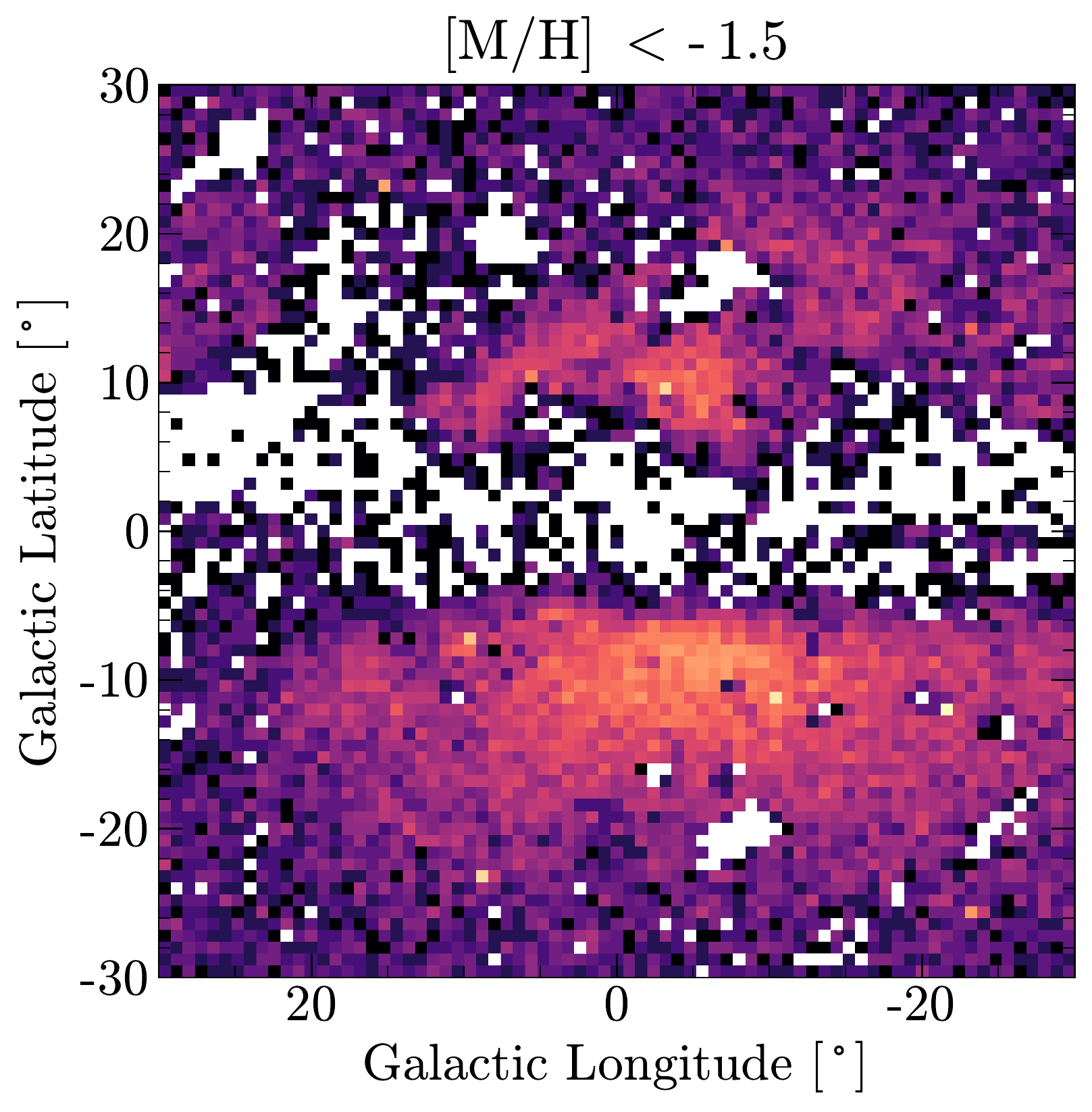}
\includegraphics[width=5.9cm]{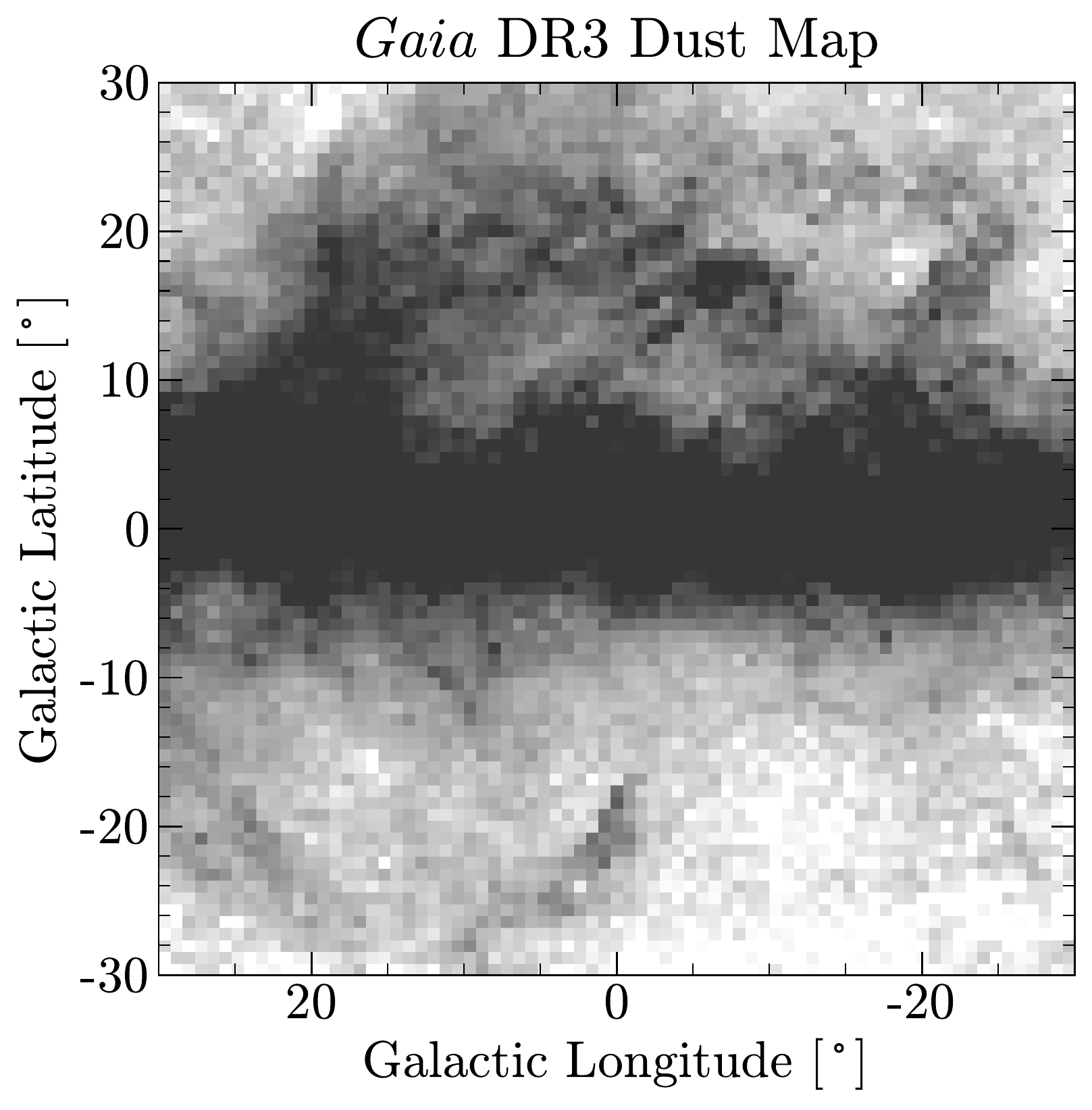}
\caption{On-sky logarithmic density distribution of stars in our sample for metal-rich (left panel) and metal-poor (middle panel) selections. The distribution of metal-poor stars appears very centrally concentrated, but the morphology is dramatically modulated by the foreground dust extinction, as illustrated in the right panel. The right panel demonstrates that the dearth of stars is highly correlated with the distribution of dust.}
\label{fig:lowMH_on_sky}
\end{figure*}

\begin{figure}[th!]
\begin{center}
\includegraphics[width=\columnwidth]{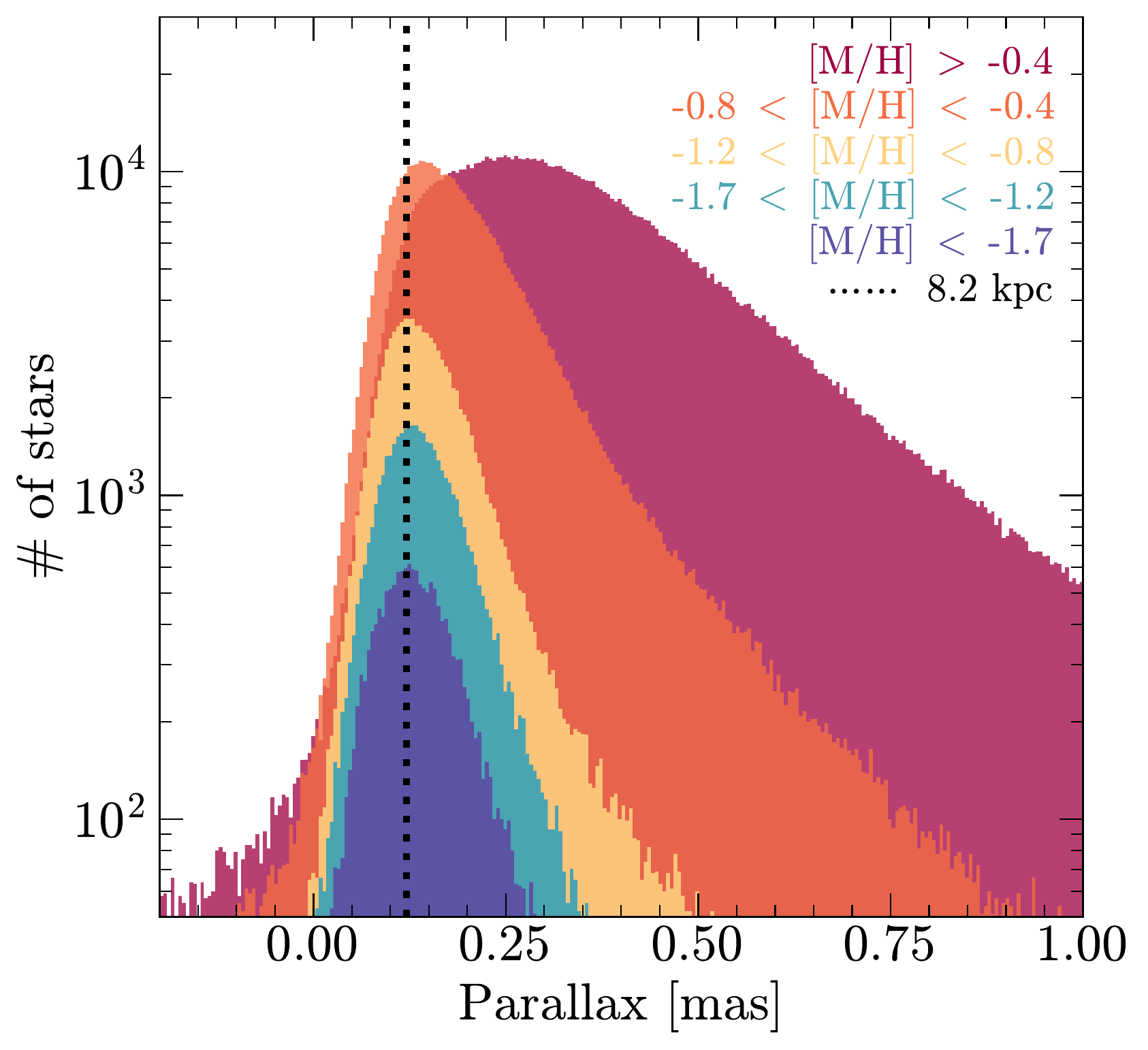}
\end{center}
\caption{Parallax distribution of the different sub-samples, selected by \MH, taking $\varpi$ as a proxy for distance. It is apparent that the metal-poor samples are centered around the expected parallax of the Galactic center (dotted black line; \citealt{GravityCollaboration2021b}), i.e. they are very concentrated near the Galactic Center (GC), increasingly so towards lower [M/H].}
\label{fig:parallax_distribution}
\end{figure}

Figure~\ref{fig:lowMH_on_sky} compares the on-sky density distribution of metal-rich stars with $\MH > -0.4$ (left panel) to metal-poor ones with $\MH < -1.5$ (middle panel).  The metal-poor stars exhibit a striking central concentration relative to the metal-rich sample, suggesting a spheroidal population in the inner Galaxy.  
%and $\hat{R}_{\text{GC}}<5$~kpc; to determine the approximate Galactocentric distance $\hat{R}_{\text{GC}}$ we have assumed that the line-of-sight distance $D_{los}$ can be by the inverse parallax, limited at $D_{los}^{max}=10~kpc$. 
The projected sky density of metal-poor sample members in Figure \ref{fig:lowMH_on_sky} (middle panel) is dramatically altered by the high dust extinction towards the GC (right panel). 
%We did require $G_{\text{BP}}<15.5$ in the selection to ensure good \MH estimates. The right panel, presenting a superposition of the stellar map and dust map (gray), shows how well and detailed 
The fine-scale structure in the projected $\MH < -1.5$ stellar distribution can mostly be explained by dust extinction, as the right-hand panel of Figure \ref{fig:lowMH_on_sky} shows. This is expected as we require $G_{\text{BP}}<15.5$ for sample selection.

We do not have precise parallaxes for the entire sample, which complicates the analysis of the spatial structure. For the overall sample, the median parallax precision is
$\varpi / \sigma_\varpi \sim 12$, with 10\% of the sample having $\varpi / \sigma_\varpi \lesssim 4$, and 1\% of the sample with negative parallaxes. For the metal-poor sample members with $\MH<-1.5$, which turn out to be more distant than the sample mean, the median precision is only 
$\varpi / \sigma_\varpi \sim 5$. Therefore, the most robust way to present the distance distribution of objects in a directional cone may be to consider their parallaxes. Given that the objects of particular interest are at $\sim 8$~kpc distance, it matters that we apply a Gaia parallax zero-point correction, for which we adopt $0.02$~mas \citep{parallax_zero_point}.

Figure~\ref{fig:parallax_distribution} shows $n_*\bigl ( \varpi~|~\MH \bigr )$ in five \MH -bins, with the minimal distance (maximal parallax) of 1~kpc at the right edge, and the parallax expected for the Galactic Center indicated by the dotted line \citep[$1/8.2$\,kpc, or 0.12~mas; ][]{GravityCollaboration2021b}. This figure shows that the distance distribution of giants towards the Galactic center depends dramatically on \MH . Metal-rich stars ($\MH >-0.4$) are distributed throughout the disk with heliocentric distances $D_\odot=1-5$~kpc (or 0.2-1.0~mas). In contrast, metal-poor populations become increasingly more concentrated toward the Galactic Center (GC). 
Indeed, the most metal-poor stars are predominantly clustered around the GC's parallax, revealing that there is a centrally concentrated very metal-poor population in the Milky Way.

Figures~\ref{fig:lowMH_on_sky} and \ref{fig:parallax_distribution} qualitatively confirm the picture of a metal-poor population that is very concentrated toward the Galactic center. The fact that there are almost no $\MH < -1.5$ stars at very low latitudes shows that almost all lie behind the strong dust extinction 2-3~kpc inward of the Sun. Rigorous modelling of these stars' spatial distribution of these stars density distribution \citep[e.g.][]{Rix2021} is beyond the scope of this paper, with the severe dust extinction and crowding effects presenting formidable challenges. However, we can model the parallax distribution (Fig.~\ref{fig:parallax_distribution}) in the central regions less affected by dust to quantify the radial extent of this metal-poor population. 

We assume that the 3D distribution is a spherical Gaussian as a function of Galactocentric radius $R$ $\rho_{Gauss}(R,\sigma_{R_{\text{GC}}})$, with a width of $\sigma_{R_{\text{GC}}}$. As long as we can see to the far side of the Galactic center (see Fig.~\ref{fig:parallax_distribution}), we should expect the parallax ($\varpi$) distribution in a cone of $d\Omega$ in the direction $(l,b)$ to be
\begin{equation}
    n(\varpi)\, = \, d\Omega\, d\varpi\, \varpi^{-4}\, \rho_{Gauss}\bigl (R( \varpi,l,b),\sigma_{R_{\text{GC}}}\bigr ).
\end{equation}
Phrasing this model in terms of the $\varpi$ allows the model to deal with vanishing (or even negative) observed parallaxes.
We find that the maximum likelihood of the observed parallaxes (and their uncertainties) for the $\MH<-1.5$ stars implies an extent of $\sigma_{R_{\text{GC}}}\sim 2.7$~kpc. Interestingly, this radius corresponds to an angle of $18^\circ$ at the distance of the Galactic center, which seems very plausible in light of the projected density distribution (Figure~\ref{fig:lowMH_on_sky}; middle panel).

\subsection{The \MH -Distribution in the Inner Galaxy}\label{sec:MH-dist}

\begin{figure}%[h]
\includegraphics[width=\columnwidth]{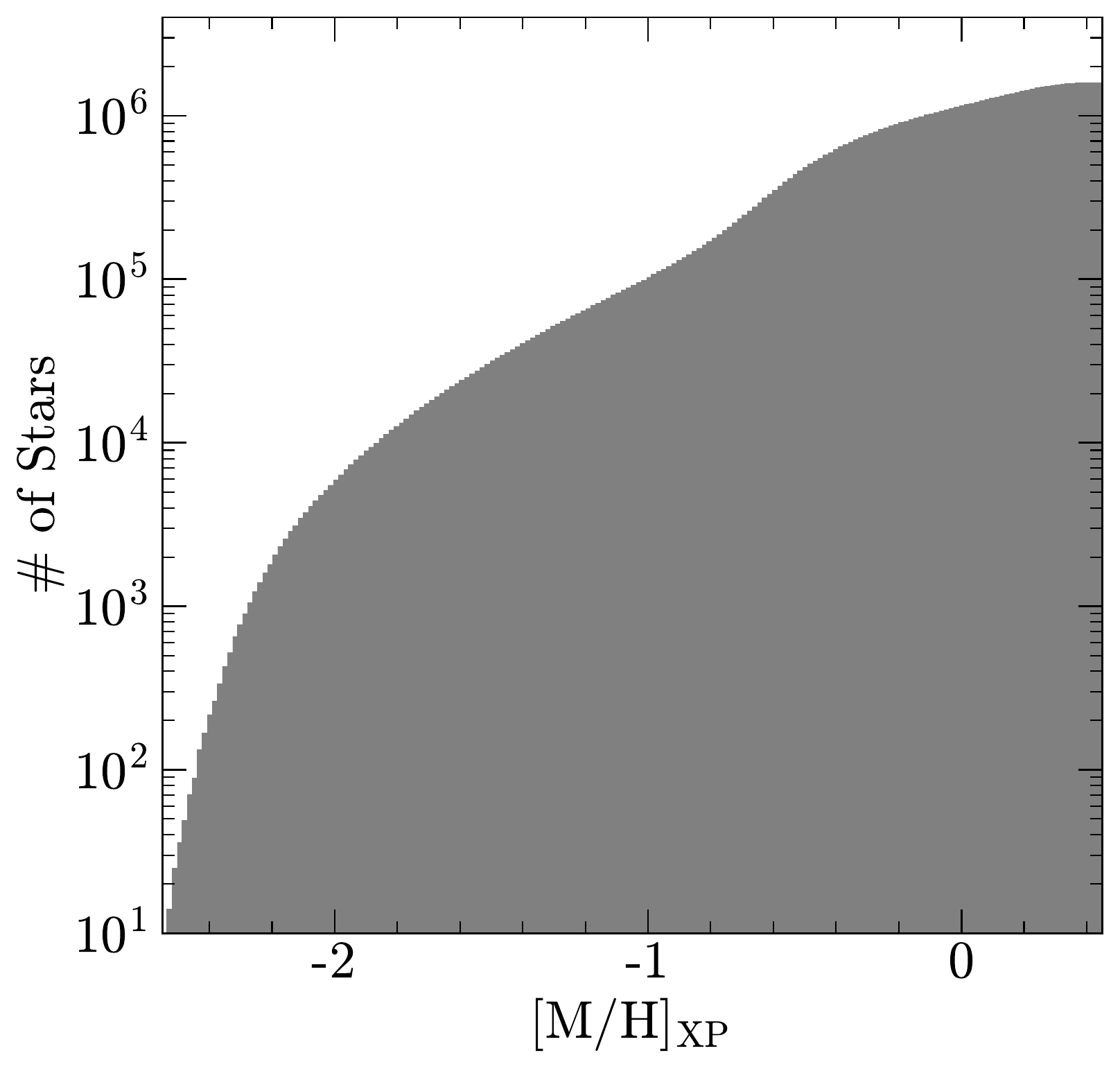}
\includegraphics[width=\columnwidth]{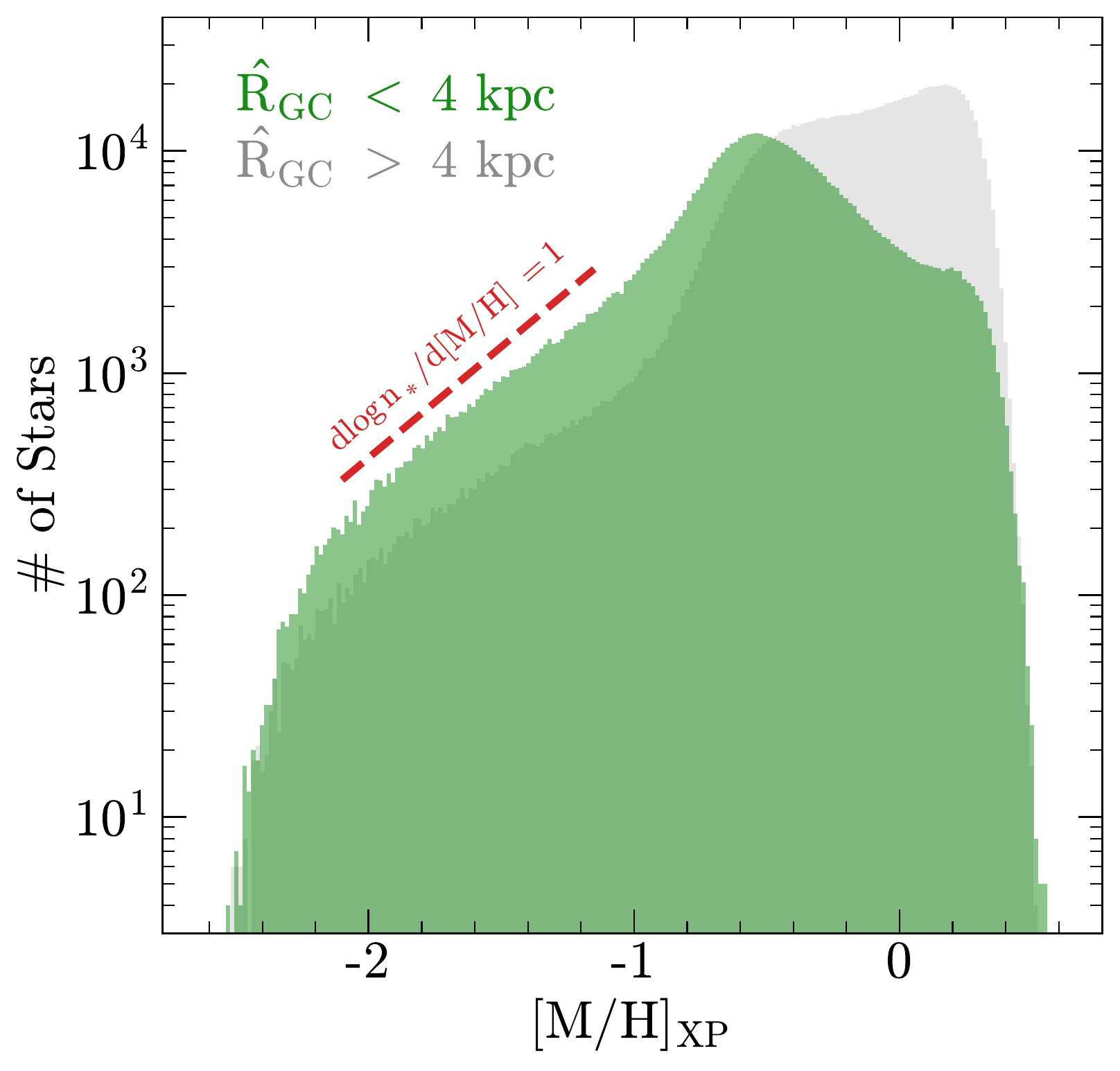} 
\caption{Top: Cumulative distribution of metallicities \MH\ for the entire sample of $\sim 1.7$ million objects, estimated from XP spectra and broad-band photometry via XGBoost prediction trained on the APOGEE DR17: there are $\sim 5,000$ stars with $\MH<-2$, $\sim 20,000$ stars with $\MH < -1.5$, and nearly 100,000 stars with $\MH < -1$. Bottom: Distribution of [M/H] for our sample, split into sources with presumed Galactocentric radii $R_{\text{GC}}<4$~kpc (in green), which are the focus of our analysis, and those at larger $R_{\text{GC}}$. The (green) distribution shows a steep rise to $\MH \sim -2$, then a power-law of slope $d(\log{n_*})/d\MH \approx 1$ to $\MH \sim -1$, and finally a yet again steeper rise to $\MH \sim -0.6$. A possible interpretation of these slopes in the context of a simple chemical evolution model is given in Section~\ref{sec:MH-dist}. 
}
\label{fig:MHdistribution}
\label{fig:cumulative_distribution}
\end{figure}

We now turn our attention to the \MH -distribution, $n_*(\MH )$, in the inner Galaxy, which is illustrated in Figure~\ref{fig:MHdistribution}. The first remarkable feature of the distribution is the sheer number of objects, most apparent from the cumulative distribution in Figure \ref{fig:cumulative_distribution}: there are $>4,000$ stars with an estimated $\MH<-2$, $\sim 18,000$ with $\MH < -1.5$, and almost 100,000 stars with $\MH < -1$, i.e. below the abundance floor of the old, $\alpha$-enhanced disk. This is an order of magnitude more objects than previously published metal-poor samples ($\MH < -1$) of the inner Galaxy \citep[e.g.,][]{Arentsen2020a,Arentsen2020b}.

A rough estimate shows that this corresponds to $\gtrsim 5\times 10^7$M$_\odot$ in stars at $\MH < -1.5$ in the inner few kpc of the Galaxy. Taking the nearest globular cluster M4 as a template for an old, $\MH < -1$ population, we find that there are about 35 giants with $M_G<0.5$ in the Gaia DR3 catalog beyond M4's half mass radius of $6^\prime$ \citep[][where the Gaia catalog should be approximately complete]{Richer2004}. Given M4's total mass of $7\times 10^4$M$_\odot$ \citep{Marks2010}, this implies one giant (at $M_G<0.5$) per 800\,M$_\odot$ of total stellar population mass. At face value, this implies $M_{tot}\bigl(\MH < -1.5\bigr) \sim 2\times 10^7$\,M$_\odot$. But this does not account for the incompleteness of XP spectra in very crowded fields, not for the far-reaching effects of dust extinction. A conservative estimate of these effects is a factor of 2.5, leading us to a lower limit on the mass of $M_{tot}\bigl (\MH < -1.5\bigr ) \gtrsim 5\times 10^7$M$_\odot$.

We have already presented two arguments that the low-\MH\ portion of $n_*(\MH )$ is not severely contaminated by spurious \MH\ estimates from some of the far more numerous stars of higher (true) metallicity: the validation with APOGEE and the parallax, or distance, distribution, which is most peaked for the most metal-poor objects.  The \MH\ distribution itself provides additional evidence. It rises very steeply from $\MH =-2.5$ to $-2.2$. The distribution then follows a power law of slope $d(\log{n_*})/d\MH \approx 1$ over a wide range of metallicities to $\MH\approx -0.9$, where it steepens towards a peak at $\MH=-0.6$, beyond which $n_*(\MH )$ starts dropping towards the highest metallicities present, $\MH\approx 0.5$. 

The bottom panel of Figure~\ref{fig:MHdistribution} shows the metallicity distribution with stars of an estimated Galactocentric distance $\hat{R}_{\text{GC}} < 4$~kpc in green, and the stars of the intervening disk in gray. As giants, in particular red clump stars, are of comparable luminosity and color over a wide range of metallicities, the \emph{differential} selection effects across different \MH~ should be modest. Hence, the \MH -distribution of stars $R_{\text{GC}}<4$~kpc in Figure~\ref{fig:MHdistribution} should be relatively unbiased. The \MH~ distribution of course varies with position, in the Galaxy (both $|z|$ and $R_{\text{GC}}$), as Figure~\ref{fig:MHdistribution} illustrates.

The decrease in $n_*(\MH )$ for stars at the metal-rich end ($\MH > 0$) with $R_{\text{GC}}<4$~kpc (green) may seem surprising, because we know from APOGEE 
\citep[e.g.][]{Eilers2022} that the inner Galaxy is teeming with high-metallicity stars. This decrease in $n_*(\MH )$ is straightforwardly explained by Figure \ref{fig:lowMH_on_sky}, which shows that in the inner Galaxy, we are almost completely lacking sample members with $|Z|<700$~pc; any thinner population will inevitably be absent from our sample due to dust extinction.  
The APOGEE sample has a different MDF in part because its 2MASS-selected stars better sample the low-$|Z|$, high-\MH\ population.

The bulk of the metal-poor population, below the minimal \MH\ of the old disk at $\MH\sim  -1$, follows $d(\log{n_*})/d\MH \approx 1$: the cumulative number of stars $N_*(<Z)$ below a (linear) metallicity $Z$ grows linearly with $Z$. This seems to be a natural slope for $n_*(\MH )$ in the early phases of a self-enriching, \insitu\ system, as shown by the following argument, which builds on the models of \cite{Weinberg2017}.  

Under fairly general conditions, the evolution of the ISM metallicity in a simple, fully mixed system can be described by
\begin{equation}
    \frac{dM_Z}{dt}=y_Z~\dot{M}_* - (1+\eta - r)\dot{M}_* Z,
\label{eq:dmzdt}
\end{equation}
where $M_Z$ is the mass in metals, $y_Z$ is the IMF-averaged yield, $\dot{M}_*$ is the star formation rate (SFR), $Z\equiv M_Z/M_g$ is the metal mass fraction of the star-forming gas, $\eta\equiv \dot{M}_{\rm out}/\dot{M}_* $ is the mass loading factor of a gaseous outflow, and $r \approx 0.4$ represents the recycling of metals that are formed into stars but quickly returned when the stars die. Equation~(\ref{eq:dmzdt}) treats enrichment as instantaneous, which should be a good approximation at low metallicities where core collapse supernovae and massive star winds are the dominant sources.  It also assumes that outflows have the same metallicity as the ambient ISM; if outflows instead consist of a fraction $f$ of the supernova ejecta plus entrained ISM, then the effect is to reduce $y_Z$ by a factor $(1-f)$ so that it reflects only metals retained by the ISM.  Importantly for the case at hand, if $Z$ is well below the equilibrium metallicity $Z_{\rm eq}\equiv y_Z / (1+\eta-r)$ then the entire second term of equation~(\ref{eq:dmzdt}) can be neglected. 

In this low metallicity limit, equation~(\ref{eq:dmzdt}) is simply $\dot{M}_Z=y_Z \dot{M}_*$, and for a metallicity-independent yield, its time integral implies $M_Z = y_Z M_*$ and thus $Z = y_Z M_*/M_g$.  The log-slope of the MDF is
\begin{equation}
   \frac{d\log{n_*}}{d\MH}~=~
    \frac{d\log{M_*}}{d\log{Z}} ~=~
    \frac{Z}{M_*}\cdot\frac{\dot{M}_*}{\dot{Z}} ~=~
    \frac{y_Z}{\dot{Z}\tau_*},
   \label{eq:enrichment1}
\end{equation}
where the last equality introduces the star formation efficiency (SFE) timescale $\tau_* \equiv M_g/\dot{M}_*$.    Using 
\begin{equation}\label{eq:zdot}
    \dot{Z} = {\frac{d}{dt}}(M_Z/M_g) = y_Z \frac{\dot{M}_*}{M_g} - y_Z \frac{M_*\dot{M}_g}{M_g^2}
\end{equation}
gives the end result
\begin{equation}
   \frac{d\log{n_*}}{d\MH}~=~
   \left(\frac{\dot{Z}\tau_*}{y_Z}\right)^{-1} ~=~
   \left[ 1 - \left(\frac{M_*}{M_g}\right)
    \left(\frac{\tau_*\dot{M}_g}{M_g}\right)\right]^{-1}~.
   \label{eq:enrichment}
\end{equation}
The combination $(\tau_*\dot{M}_g/M_g)$ corresponds to the fractional change of the gas reservoir mass over one SFE timescale.

Equation~(\ref{eq:enrichment}) shows than an MDF log-slope $\approx 1$ is a generic result in the low metallicity regime ($Z \ll Z_{\rm eq}$), arising whenever the gas mass is constant ($\dot{M}_g=0$) or more generally when the star-to-gas ratio $M_*/M_g$ is small enough that the second term in Equation~(\ref{eq:zdot}) can be neglected.  Figure~\ref{fig:MHdistribution} exhibits this generic slope over the range $-2<\MH<-1$.  The steeper slope at $\MH<-2$ could plausibly arise because the gas reservoir is small but rapidly growing, with $\tau_* \dot{M}_g/M_g > 1$.  The steeper slope at $-1 < \MH < -0.6$ may reflect rapid gas accretion (high $\dot{M}_g/M_g$) coinciding with the onset of the old $\alpha$-enhanced disk \citep[e.g.][]{Belokurov2022,XiangRix2022,Conroy2022}.  At still higher metallicities our neglect of sink terms in equation~(\ref{eq:dmzdt}) becomes a poor approximation.

\subsection{The Orbit Distribution of low-\MH\ Stars in the Inner Galaxy}

We now turn to the orbit distribution of the sample at hand (see Section~\ref{sec:orbit_calculation}), which can tell us which of the metal-poor stars \emph{selected} in the inner Galaxy remain confined to the central regions of the Milky Way. The orbit distribution can tell us which stars are just passers-through near their pericenter, but on orbits that take them into the outer halo. In particular, we might expect to see the pericenter members of accreted halo components like GSE, with stars on highly eccentric orbits that take them to $R_{\text{apo}}>10$~kpc.  We can also delineate how early -- or how metal-poor, if we use abundances to estimate relative stellar ages --  the population develops net rotation and how rapidly the kinematics change with \MH~ towards near-circular motion in the plane of the Galaxy \citep[see][]{Arentsen2020a, Belokurov2022,Conroy2022}.

\begin{figure}[t!]
\begin{center}
\includegraphics[width=\columnwidth]{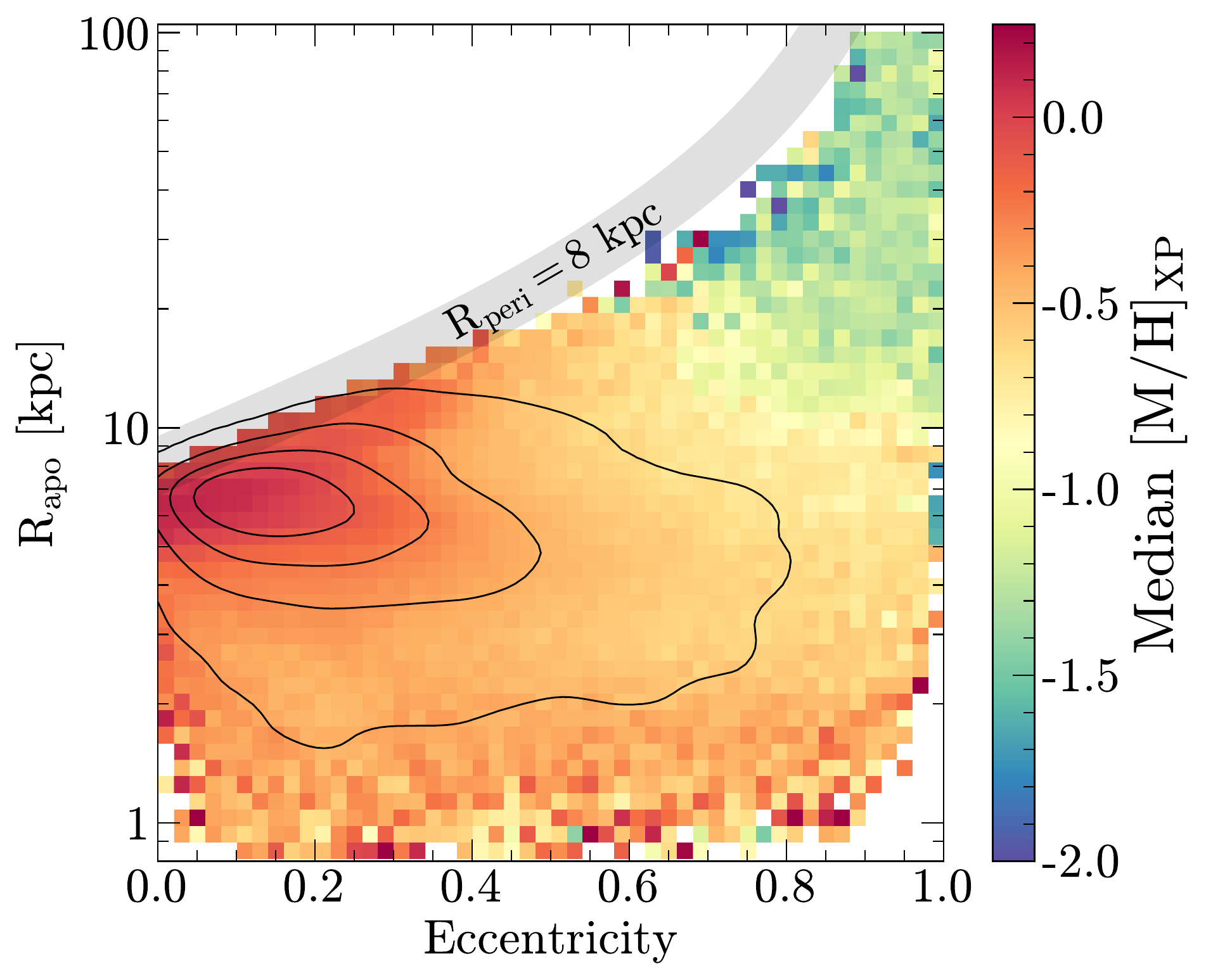}
\includegraphics[width=\columnwidth]{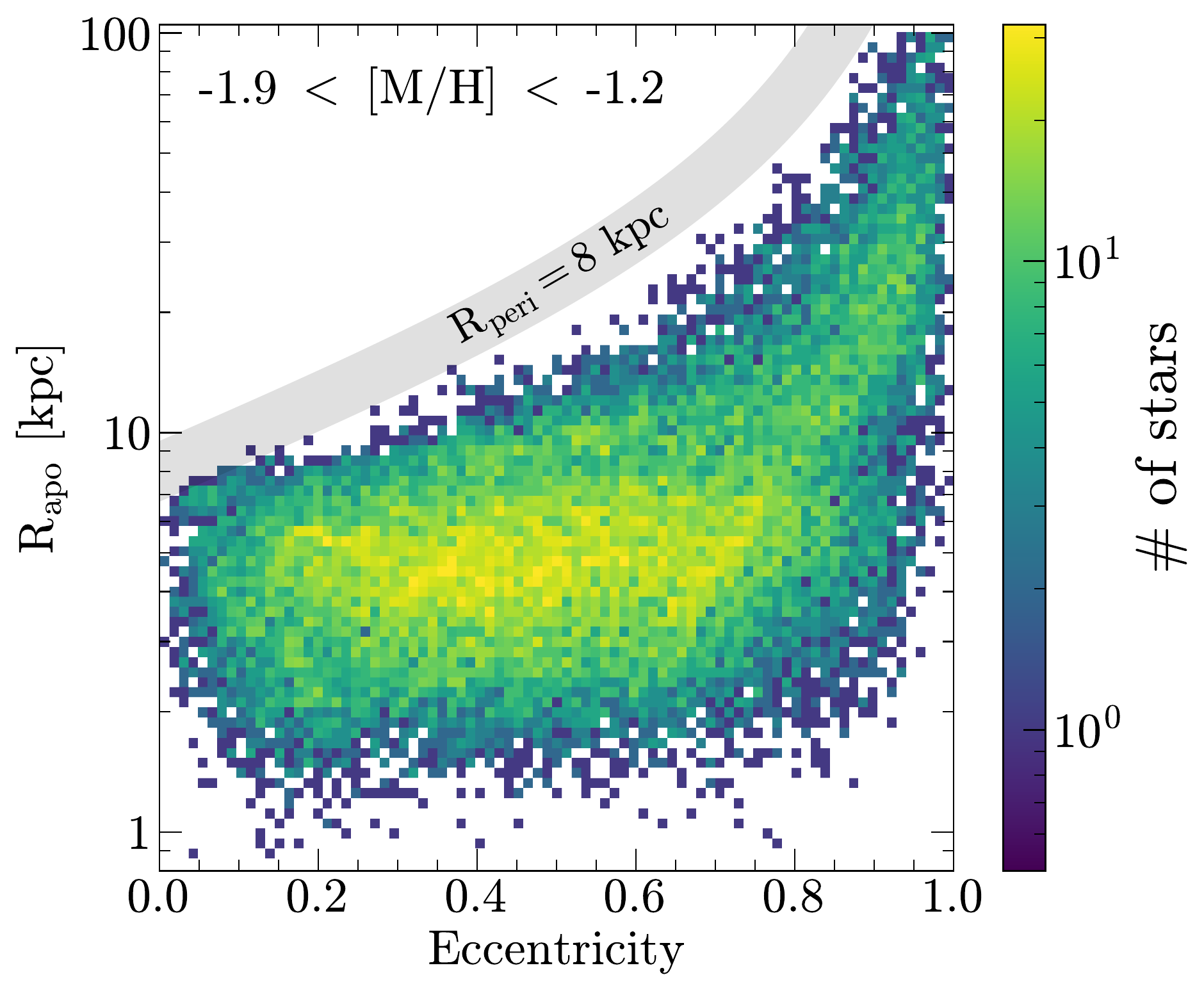}
\end{center}
\caption{Top: $R_{\text{apo}}$ \emph{vs.} eccentricity space colored by the median XP metallicity. We overlay contours of linear density using a Gaussian kernel density estimate. Bottom: Distribution of the $-1.9<\MH<-1.2$ sub-sample in $R_{\text{apo}}$ \emph{vs.} eccentricity  space. This figure shows that most of the low-metallicity sample members have orbits that remain confined to the inner Galaxy ($R_{\text{GC}}<5$ kpc), with a broad eccentricity distribution. At high eccentricities, there are many stars with large apocenters, potentially representing a population of accreted stars on highly radial orbits.}
\label{fig:ecc_Rapo}
\end{figure}

We start by considering the orbit distribution, $n_*(e,R_{\text{apo}})$,  in terms of the orbital eccentricity, $e$ and the apocenter $R_{\text{apo}}$. We initially focus on the \MH -range $-1.9<\MH<-1.2$ since this encompasses the bulk of known GSE stars. This distribution of 28,000 stars is shown in Figure~\ref{fig:ecc_Rapo}. The density in the $(e,R_{\text{apo}})$-plane is affected and limited by two experimental aspects. First, our initial Gaia query selects almost exclusively stars with $R_{\text{GC}} < 8$~kpc, and hence all stars whose pericenter distance exceeds $8$~kpc are excluded; this boundary is indicated by the thick gray line. Except at $e<0.2$ the $n_*(e,R_{\text{apo}})$ distribution falls off toward large $R_{\text{apo}}$ well before this limit.
There may be two reasons for the seeming dearth of stars with $R_{\text{apo}}<2$~kpc near the center. First, severe dust extinction simply obscures much of the central kiloparsec (Fig. \ref{fig:lowMH_on_sky}).  Second, any uncertainties in the 6D phases-space coordinates, especially uncertainties in the distance, are far more likely to increase the inferred $R_{\text{apo}}$ than decrease it, as it is a positive definite quantity. These aspects deserve careful modeling, which is beyond the scope of this paper.

The distribution in Figure \ref{fig:ecc_Rapo} can be characterized by two regimes:
%\footnote{The field of Galactic Archeology has a long and partially proud tradition of parsing continuous distributions into distinct components; in the case at hand, such a separation into two regimes seems justified, even if more exacting standards were applied.}
%DW It's a lovely footnote, but probably commenting out is wise.
most stars form a nearly flat distribution in eccentricity from $e=0.1$ to $e=0.8$, but with a narrow range in $R_{\text{apo}}$, with $\langle R_{\text{apo}}\rangle\approx 4$~kpc. This implies that most stars form an approximately isotropic distribution that stays confined to the inner Galaxy and hence will not appear in surveys of halo stars that focussed on larger $R_{\text{GC}}$ and off the Galactic plane.
The second regime is that of $e \ge 0.75$ and $R_{\text{apo}} \gtrsim 10$~kpc. The orbits of these stars fully match the expectations of the pericenter members of the GSE stars, the accreted component that dominates the halo population between 10 and 30~kpc \citep{Helmi2018,Belokurov2018b,Naidu2021}. Indeed, recent work by \citet{Belokurovs_wrinkles} has shown quite clearly that there are likely GSE members with pericenters within 3~kpc. There is tantalizing but inconclusive evidence from Figure \ref{fig:ecc_Rapo} that the distribution of GSE stars might extend to small $R_{\text{apo}}$ and lower $e$: Figure \ref{fig:ecc_Rapo} indicates an excess of stars near $(e,R_{\text{apo}})\approx(0.7,4$~kpc). Here it would become important to disentangle GSE from other proposed accreted structures in the inner Galaxy \citep[e.g.,][]{Kruijssen2019,Horta2021}. 

%=========

% \begin{comment}
% \begin{figure*}
% \centering
% \includegraphics[width=4.5cm]{Lz_Rapo_at_vlowMH.pdf}\hfil
% \includegraphics[width=4.5cm]{Lz_Rapo_at_lowMH.pdf}\hfil 
% \includegraphics[width=4.5cm]{Lz_Rapo_at_intermMH.pdf}\hfil 
% \includegraphics[width=4.5cm]{Lz_Rapo_at_highMH.pdf}\hfil   
% %\subfloat[E]{\includegraphics[width=4.2cm]{Lz_Rapo_at_highMH.pdf}}\hfil
% \caption{The distribution of angular momentum and apocenter radii (analogous to total orbital energy) for sub-samples of increasing metallicity. As metallicity increases, the stars transition from a hot isotropic orbit distribution to a cold rotationally supported one}\label{fig:Lz_Rapo_MH}
% \end{figure*}
% \end{comment}

%=========

\begin{figure}
\centering
\includegraphics[width=\columnwidth]{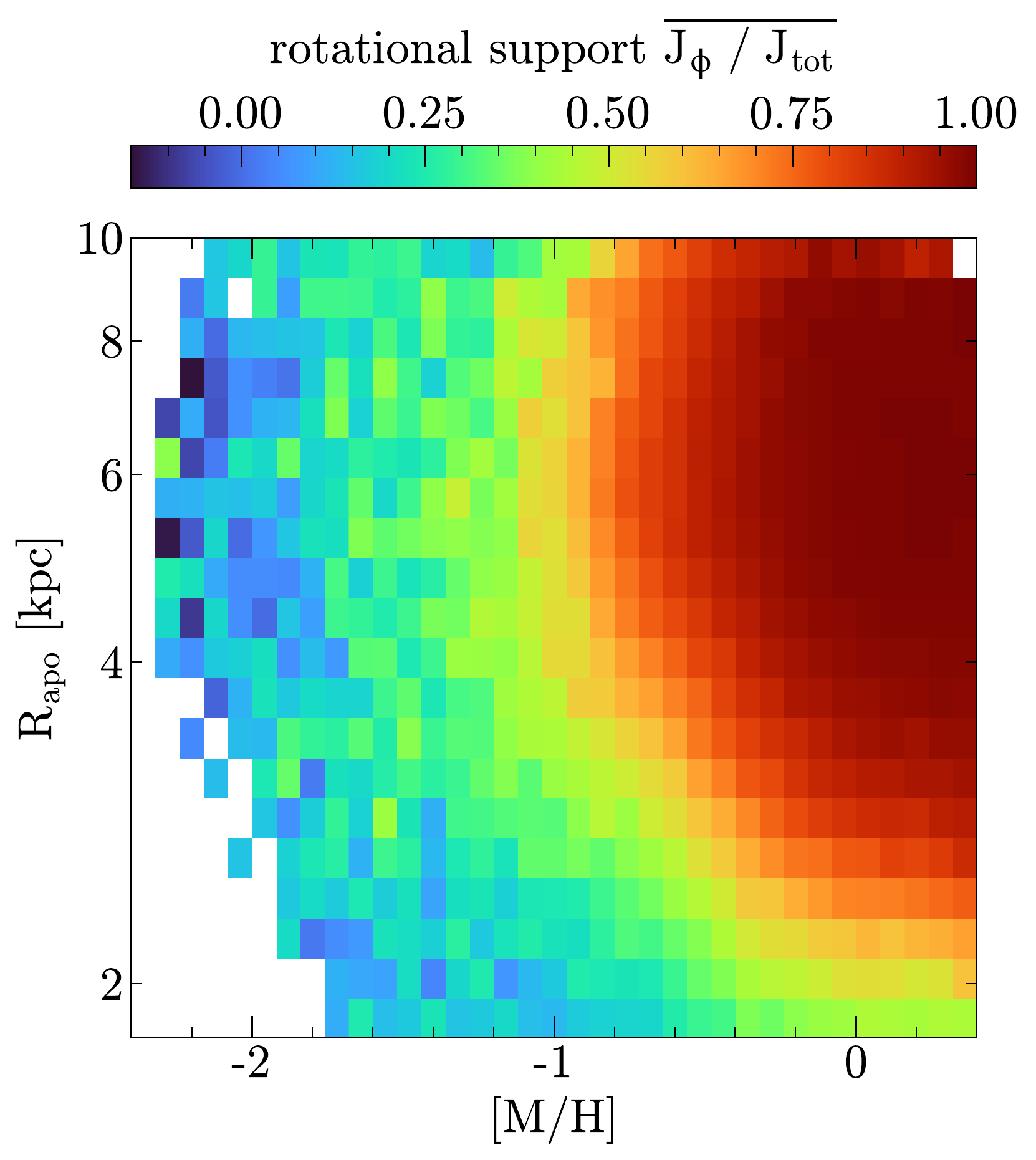}\
%hspace{0.5cm}
\includegraphics[width=\columnwidth]{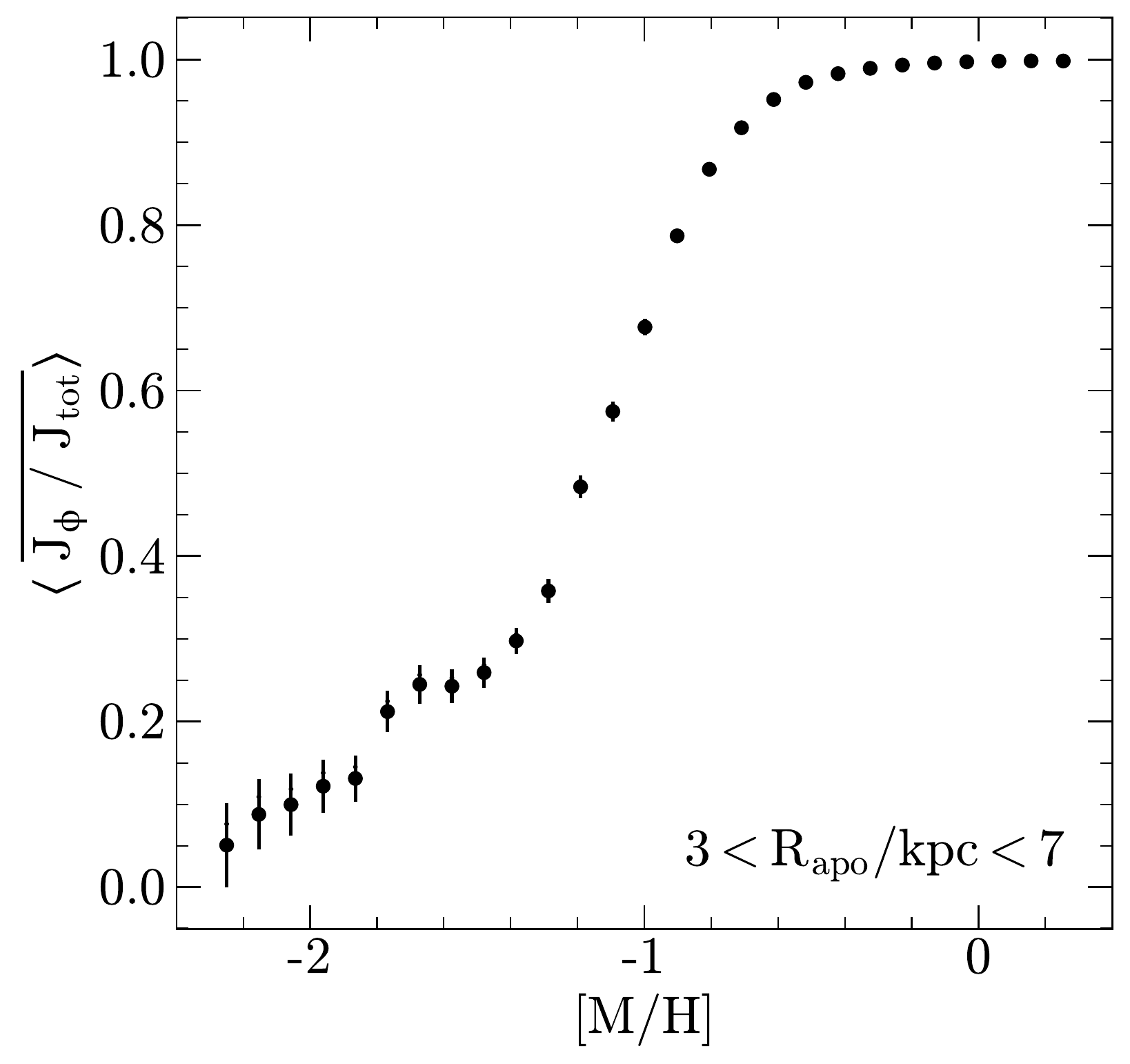}\hfill
\caption{Dependence of the orbits' level of rotation support, as a function of [M/H] and $R_{\text{apo}}$. The rotation support is defined as $\langle J_\phi/J_{\text{tot}}\rangle $, where 1 is an ensemble of circular, in-plane orbits, covering both an increase in net rotation and the approach towards circular, co-planar orbits. The top panel shows $\langle J_\phi/J_{\text{tot}}\rangle $ as a function of \MH ~and $R_{\text{apo}}$. The bottom panel shows 
$\overline{J_\phi/J_{\text{tot}}}(\MH )$, marginalized over $3~\text{kpc} <R_{\text{apo}}<7~\text{kpc}$: starting already at $\MH<-2$ there is some net $\overline{J_\phi/J_{\text{tot}}}(\MH )$, which increases to $\sim 0.3$ at $\MH=-1.3$, then steeply towards $\MH\sim -0.7$, after which it levels off in the cold rotation dominated regime.}
\label{fig:rotation_support}
\end{figure}

We now ask to what extent different metal-poor stars in the inner Galaxy have net angular momentum. In general, more metal-rich stars are generally expected and observed to have a more coherent sense of angular momentum. 
We explore this in the two panels Figure~\ref{fig:rotation_support}, which show how close, or far, stars at a given \MH\ and $R_{\text{apo}}$ are from being an ensemble of stars on circular, in-plane orbits. This is quantified via the ratio of the angular momentum, which is the azimuthal action $J_\phi$, to the total action $J_{\text{tot}}=\sqrt{J_\phi^2+J_R^2+J_z^2}$. The top panel shows that the increase of $\overline{J_\phi/J_{\text{tot}}}$ with \MH\ depends somewhat on the orbit size: At a given \MH orbits with smaller $R_{\text{apo}}$ are farther away from the coplanar quasi-circular orbits than at $R_{\text{apo}}\sim 5$~kpc. 
%We explore this via the panels of Figure~\ref{fig:Lz_Rapo_MH}, which show the distribution of different \MH -selected sub-samples in the plane of specific angular momentum, $L_z$ (or azimuthal action $J_\phi$) and apocenter distance. This shows that only the very most metal-poor stars (4.184 stars with $\MH<-1.9$) lack a distinct sense of rotation. The subsequent panels show that all the subsets of higher \MH ~also have higher mean angular momentum. These results confirm and expand what \citet{Arentsen2020a} have found.

The bottom panel of Figure~\ref{fig:rotation_support} shows $\overline{J_\phi/J_{\text{tot}}}(\MH )$, marginalized over $3 < R_{\text{apo}} / \text{kpc} < 7$: starting already at $\MH<-2$ there is some, albeit small, net $\overline{J_\phi/J_{\text{tot}}}(\MH )$, which increases gradually to $\sim 0.3$ at $\MH=-1.3$. Among more metal-rich populations, $\overline{J_\phi/J_{\text{tot}}}(\MH )$  steeply rise with rising \MH~ to $\MH\sim -0.7$, after which it levels off in the regime dominated by cold rotation. This analysis implies that some modest net rotation is present in the population well below $\MH<-1.5$, affirming and extending the findings of  \cite{Arentsen2020a}, \cite{Belokurov2022}  and \cite{Conroy2022}.
%this is different from what \citet{Belokurov2022} inferred, whose analysis implies no net rotation at $\MH=-1.5$.

\subsection{Abundance Patterns in the Inner Galaxy}

\begin{figure*}
\begin{center}
\includegraphics[width=\textwidth]{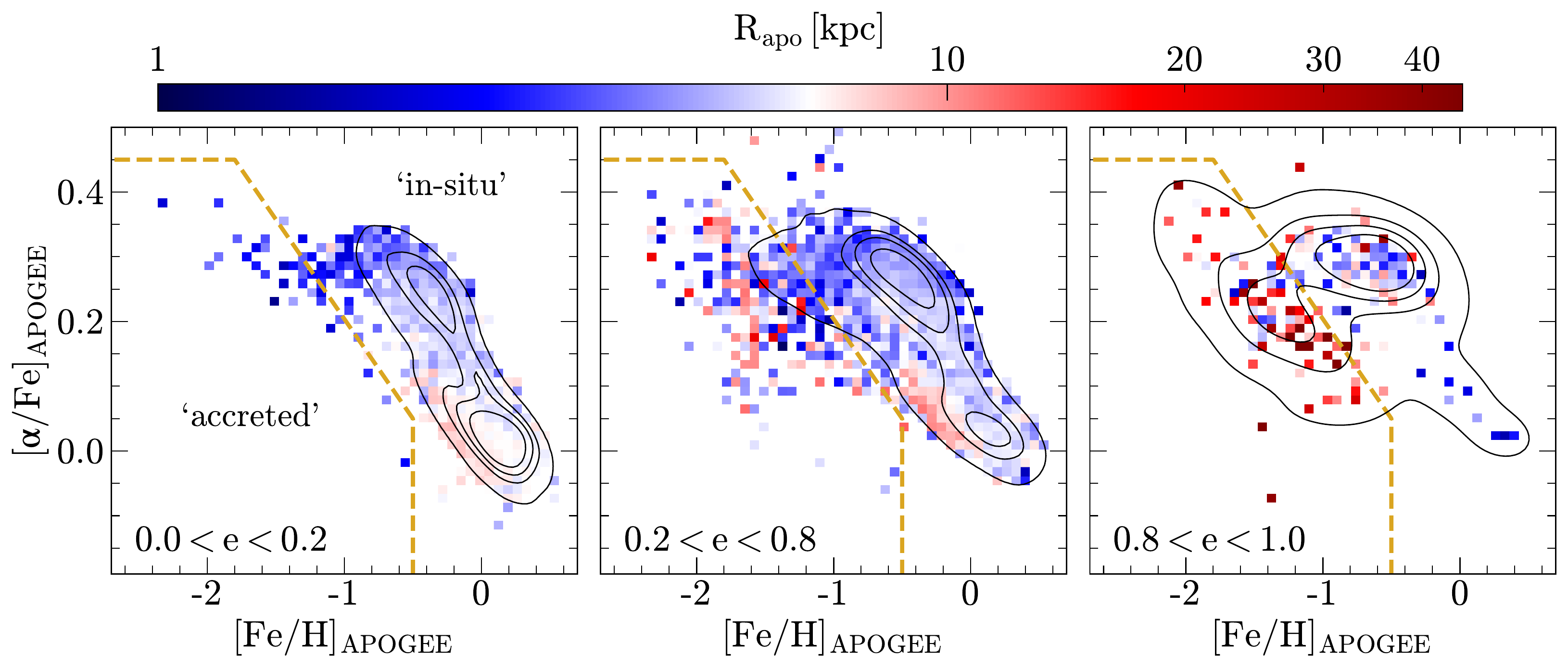}
\includegraphics[width=\textwidth]{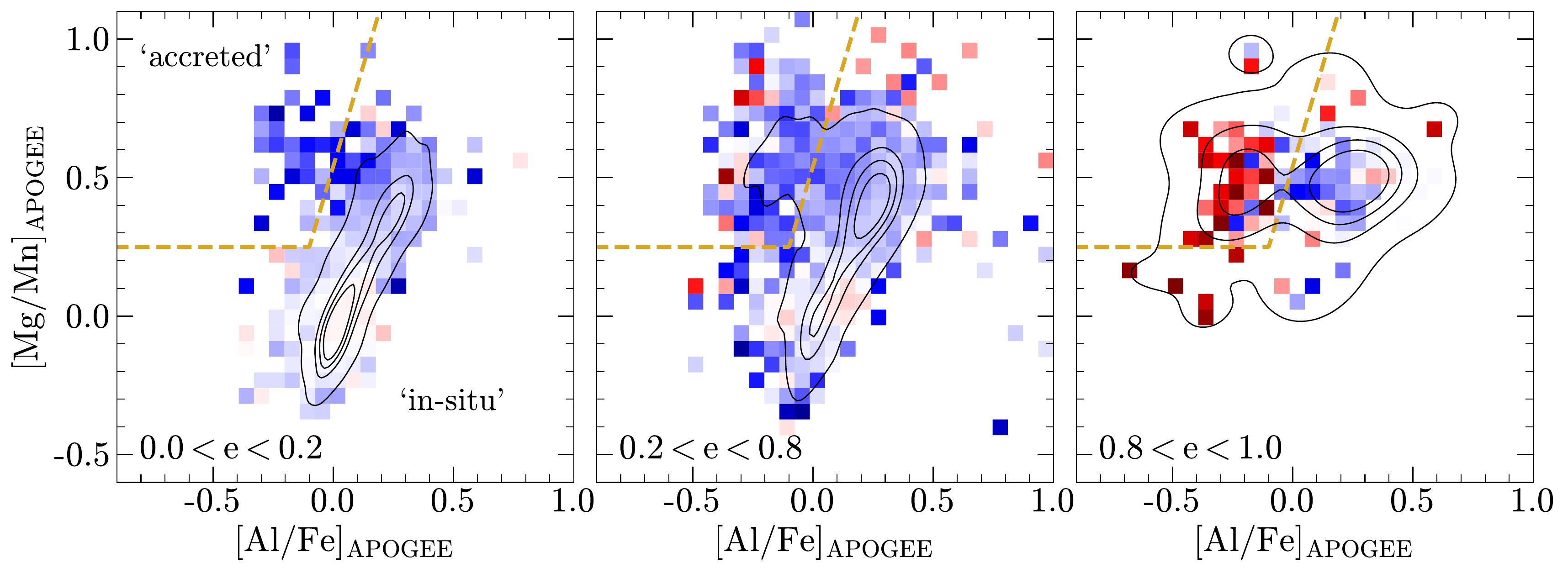}
\end{center}
\caption{Abundance diagnostics for stars in the inner Galaxy, and their relation to these stars' orbital apocenters. We show [Fe/H], [$\alpha$/Fe] (top row) and [Mg/Mn] and [Al/Fe] (bottom row) from SDSS/APOGEE (DR17) for stars  with $\varpi < 0.25$~mas. The three columns of panels show slices of increasing orbital eccentricity from left to right, with stars colored by their median $R_\text{apo}$. The color map transitions from blue to red for apocenters at the Solar Galactocentric radius; linear density contours of the underlying sample are overlaid in black. The gold dashed lines indicate boundaries that have been used for selecting `accreted' stars on the basis of their abundance patterns \citep[e.g.,][]{Hawkins2015,Das2020,Horta2021,Belokurov2022,Conroy2022}. Note, however, that the [Mg/Mn]--[Al/Fe] diagnostic has been found to be more ambiguous at low metallicity \citep[see][]{Belokurov2022,Conroy2022}. Overall, the figure shows that there is a high degree of correlation between the orbital and the abundance properties. At eccentricities $e<0.8$ most stars with $\MH<-1$ have both $R_{apo}\lesssim R_{\text{GC}}(Sun)$ and abundance patterns attributed to \insitu~ formation. Only at the highest eccentricities, $e\ge 0.8$, is there a good fraction of stars that have large apocenters and at the same time abundance patterns characteristic 
of accreted stars. Both of these latter properties are expected for members of GSE near pericenter. 
%Most stars are strongly $\alpha$-enhancement, $[\alpha / M]\gtrsim 0.25$, indicating an \insitu origin. 
%Overall, 10\% of all these stars have at most modest  $\alpha$-enhancement, $[\alpha / M]<0.15$, [\emph{ex situ}, GSE?]. Note that for eccentricities $<0.4$, only one of 73 stars has $[\alpha / M]<0.15$: all the rest is strongly $\alpha$-enhanced. 
 }
\label{fig:aM_ecc_Rapo}
\end{figure*}

We now enlist element abundances for our sample stars that overlap with APOGEE to explore which parts of the sample appear accreted from a chemical enrichment perspective (as opposed to \emph{proto-Galactic}, or \insitu ). Such abundance-based arguments for the origin of stars are based on the idea that enrichment works differently in (sub-)halos of different mass \citep[e.g.,][]{Hawkins2015,Das2020,Horta2021,Belokurov2022}:
the production of $\alpha$-elements and Al aided by high star-formation intensities in the deeper potential wells of the proto-Galaxy, compared to the shallower potential of lower-mass satellite galaxies.  However, these abundance diagnostics of the stars' formation environment may be less discriminating at lower metallicities \citep[e.g.,][]{Conroy2022}. 

We remove stars belonging to globular clusters and apply a few basic quality cuts to the APOGEE abundance data based on their uncertainties and flags. We plot the resulting chemical distributions, [X/Fe] vs. [Fe/H] with X$\, =\left\{\alpha,\,\text{Al},\,\text{Mn}\right\}$, as a function of eccentricity in Figure~\ref{fig:aM_ecc_Rapo}, , by median apocenter radius.  The presumed boundaries between \insitu~ and accreted are indicated by the yellow dashed lines.

Figure~\ref{fig:aM_ecc_Rapo} shows clearly that at modest eccentricities (left and middle columns), most stars belong to the $\alpha$-enhanced sequence, implying a proto-Galactic origin. At high eccentricities (right columns), a prominent low-$\alpha$ sequence with large apocenters appears. These stars on eccentric orbits with large apocenter appear to be remnants of the accreted GSE merger based on their chemistry and kinematics \citep[e.g.,][]{Mackereth2019,Naidu2020,Bonaca2020,Hasselquist2021,Horta2022a}. Note that among the stars of highly eccentric orbits there is a portion whose abundances point towards proto-Galactic origin (right column). Remarkably, they almost exclusively have apocenters $\lesssim 5$~kpc, very much like their cousins on less eccentric orbits, but very unlike the chemically identified GSE debris. The population of eccentric stars attributable to the proto-Galaxy has been seen before. It has been dubbed `\textit{Aurora}' by \cite{Belokurov2022} and has been explored at even lower metallicities by \cite{Conroy2022}. Our sample probes to considerably lower $R_{\text{GC}}$ than either of these works, and reveals the bulk and full extent of the metal-poor proto-Galaxy at the heart of the Milky Way. 

It is apparent from Figures~\ref{fig:ecc_Rapo} and~\ref{fig:aM_ecc_Rapo} that both orbits and abundances can help to differentiate between presumed proto-Galactic stars and stars accreted later from a lower-mass satellite. A differentiation by chemistry alone appears increasingly difficult at low \MH~\citep{Conroy2022}. But the combination of both aspects, summarized in Figure~\ref{fig:aM_ecc_Rapo}, can make such a differentiation more convincing.  

\cite{Horta2021} selected ``accreted'' stars in these APOGEE chemical spaces as evidence of a past `Heracles' merger buried deep within the MW, distinguished from GSE on the basis of a bimodal distribution in total orbital energy. This claimed galaxy bears resemblance to the previously proposed Kraken/Koala mergers \citep{Kruijssen2019,Kruijssen2020,Forbes2020}, and these names may well refer to the same event. However, \cite{Lane2022} pointed out that the apparent bimodality in orbital energy used by \cite{Horta2021} to select Heracles/Kraken/Koala might be an artifact of the spatial selection function of APOGEE. Furthermore, Heracles/Kraken/Koala and the proto-Galactic \textit{Aurora} are virtually indistinguishable in all chemical spaces \citep{Horta2022a,Naidu2022b,Myeong2022}. 

Taken together, these lines of evidence suggest that the stars attributed to  \emph{Heracles, Kraken, Koala} and \emph{Aurora}, along with the numerous metal-poor stars identified here are different orbit and metallicity regimes of one proto-Galactic stellar population, now forming the metal-poor, old heart of the Milky Way. Whereas past studies have mainly identified the eccentric tail of this population that traverses the solar neighborhood or off the Galactic plane \citep{Belokurov2022,Conroy2022,Myeong2022}, we now show more directly that the bulk of this population resides in the bulge itself, mostly within a few kpc of the Galactic center. Indeed, this matches theoretical predictions from cosmological zoom-in simulations, which predict a spheroidal distribution of \textit{proto-Galaxy} stars predominantly concentrated around the Galactic center (e.g., \citealt{El-Badry2018a}, \citealt{Wetzel2022}, Figure 13 in \citealt{Belokurov2022}). 

The picture presented here appears to obviate the need for an \textit{ex-situ} merger like Heracles/Kraken/Koala, at least based on the orbits and chemistry of field stars. This does not necessarily imply that these mergers did not occur; rather, it emphasizes the -- conceptual and practical -- ambiguity between \insitu~ and \textit{accreted} stars during the earliest phases of the Milky Way, which led us to use the collective term proto-Galaxy. This view is supported by cosmological zoom-in simulations, which affirm the difficulty in distinguishing accreted galaxies from the MW proto-galaxy at early times \citep[e.g.,][]{Renaud2021a,Renaud2021b,Orkney2022}. 

\section{Summary, Discussion and Outlook}\label{sec:discussion}

We have presented an exploration of the metal-poor stellar population in the inner Galaxy ($R_{\text{GC}}\lesssim 5$~kpc), drawing on the newly available low-resolution XP spectroscopy from Gaia's DR3. The primary goal was to see whether there is an extensive ancient and metal-poor stellar population at the ``heart'' of the Milky Way. We wanted to learn whether the bulk of these stars are a tightly bound proto-Galactic population, rather than stars accreted from more distant satellites, whose location in the inner Galaxy just reflects the pericenter phase of orbits that take them out to the ``classical'' halo at $R_{\text{GC}}\gtrsim R_\odot$. As simulations imply that the earliest heart of galaxies often arises from the high-redshift coalescence of clumps with comparable mass, we prefer the term \emph{proto-Galactic} for the resulting population, rather than \insitu, a term that singles out one of these clumps even if it has only modestly larger mass than other nearby clumps.

The practical foundations of our analysis were estimates of \MH\ of $\sim 2$~million giant stars within 30$^\circ$ of the Galactic center, at least 1~kpc from us and sufficiently bright for good S/N in the blue part of the XP spectra, $\GBP\le 15.5$. We converted the XP spectral information into a set of narrow-band fluxes in filters that are known to be highly diagnostic of \MH\ in cool stars. We combined this synthetic photometry with ALLWISE photometry and then used them -- along with the XP spectral coefficients -- for a data-driven \MH\ prediction, trained on the APOGEE DR17 data using XGBoost. We found these \MH\ estimates to be precise to $\sim$0.1 dex. We also found them robust and pure enough to identify metal-poor samples of stars in the inner Galaxy $(\MH <-1)$, despite the presence of a vastly dominant population of more metal-rich stars in the inner Galaxy. For most of these stars, radial velocities exist from Gaia RVS, allowing the estimates of their orbits. 

On this basis, we identified metal-poor samples of stars towards the inner Galaxy that are about two orders of magnitude larger than published ones: $>4,000$ stars with $\MH <-2$, $\sim 20,000$ stars with $\MH <-1.5$, and $\sim 70,000$ stars with $\MH <-1$.
The most metal-poor tail of the metallicity distribution extends to $\MH\approx -2.5$. Whether this constitutes a genuine metallicity floor or reflects that the APOGEE training set for XGBoost only extended to $\MH=-2.5$ requires follow-up data. 

These samples reveal a large metal-poor population ($\MH < -1.5$) in the inner few kpc of our Galaxy. Three lines of argument imply that this population is centrally concentrated: first, the parallax distribution implies that most stars are within 5~kpc of the Galactic center (Figure~\ref{fig:parallax_distribution}). If we model the spatial density distribution of stars with $\MH<-1.5$ as a Gaussian centered on the Galactic center, with an extent of $\sigma_{R_{\text{GC}}}$, this parallax distribution implies $\sigma_{R_{\text{GC}}}\sim 2.7$~kpc. Second, the number density projected onto the sky shows a centrally concentrated distribution (albeit severely modulated by intervening dust, Figure~\ref{fig:lowMH_on_sky}), which is qualitatively consistent with $\sigma_{R_{\text{GC}}}\sim 2.7$~kpc. Third, their orbits show that most stars have \emph{apocenters} of less than 5~kpc (Figure~\ref{fig:ecc_Rapo}). The latter property implies that most of this population could not have been found in many of the previous surveys that focused on
$R_{\text{GC}} > 5$~kpc.  At the same time, our analysis showed that a minority (but still a large number) of stars with $\MH < -1.2$ currently in the inner Galaxy are on highly eccentric orbits that can take them to $>10$~kpc. These are presumably members of the accreted halo, found here near the pericenter of their orbits.

Using the members of this sample that have detailed abundances from SDSS (DR17), in particular [$\alpha$/Fe], [Al/Fe] and [Mn/Fe], we explored which of these stars have abundance patterns attributable to a proto-Galactic (or, \insitu ) vs. accreted origin. We found that almost all stars that remain tightly bound within the inner Galaxy ($R_{\text{apo}}<R_{\text{GC}}(\text{Sun})$) have abundances that point towards a proto-Galactic origin. Stars on highly eccentric orbits with $R_{\text{apo}}>10$~kpc show abundances typical of accreted stars and are presumed pericenter members of GSE. 

We quantified how much net rotation the various mono-abundance populations show and how much they resemble disk-like orbits. We did this by considering the mean ratio of the angular momentum to the total action, $\overline{J_\phi/J_{\text{tot}}}(\MH )$. 
We found that only the most metal-poor tail $\MH<-2$ of this population shows no rotation, $\overline{J_\phi/J_{\text{tot}}}(\MH <-2)\approx 0$.  Populations with higher $\MH$ also have higher $\overline{J_\phi/J_{\text{tot}}}(\MH )$: at $\MH=-1.5$, $\overline{J_\phi/J_{\text{tot}}}$ has reached 0.25, then continuously increasing to $\overline{J_\phi/J_{\text{tot}}}\approx0.9$ in $\MH=-0.8$.

All of this information fits a picture in which this metal-poor heart of the Milky Way constitutes the most ancient \textit{proto-Galactic} component of our Galaxy. 
\begin{itemize}
    \item Much of this population is confined to $R_{\text{GC}}<R_\odot$, or tightly bound at the bottom of our Galaxy's potential well, arguing against an origin through accretion from a once-distant satellite.
    \item The vast majority of stars with $\MH < -1.5$ are distinctly $\alpha$-enhanced, arguing for rapid enrichment in a deep potential well.
    \item The stellar mass we see at $\MH <-1.5$ within 5~kpc corresponds to $\sim 5\times10^7\mathrm{M}_\odot$, but manifest dust obscuration implies a large correction that suggests that the stellar mass of this metal-poor heart of the Milky Way is high, $\mathrm{M}_*\gtrsim 10^8\mathrm{M}_\odot$. Such high masses of strongly $\alpha$-enhanced stars at such low \MH\ are most likely to occur at the center of a (eventually) quite massive halo, not at a low-mass satellite.
    \item We have mapped the degree of (prograde) ordered azimuthal motion, $\overline{J_\phi/J_{\text{tot}}}(\MH )$ as function of \MH . At $\MH\sim -2$ it is approximately zero, and increases continuously when considering sets of stars with increasing \MH. Such a relation is expected if star formation  occurs -- as time goes on -- in gas that has increasingly settled into a disk, as recent simulations of disk galaxy formation show \citep{Gurvich2022}.
   
    \item Much of this central component has \MH\ well below that of the centrally concentrated, $\alpha$-enhanced, thick disk, whose oldest members (at $\MH\sim -1$) appear to be $\approx 12.5$~Gyrs old \citep{XiangRix2022}. If much of this metal-poor heart of the Milky Way chemically, and therefore temporally, predates the oldest $\alpha$-enhanced disk, it must have formed at $\tau\gtrsim 12.5$~Gyrs, corresponding to $z\gtrsim5$. This stellar population at the heart of the Milky Way should be almost exclusively ancient.
\end{itemize}

The results presented here are by no means a \emph{new distinct} stellar component of the Milky Way. The stars in our sample seem to constitute the tightly bound part -- and bulk -- of a proto-Galactic spheroid. The distribution tail of stars on somewhat more extended orbits has already been recognized in recent work as \insitu ~halo \citep{Belokurov2022,Conroy2022}. But our results significantly flesh out the existing picture by showing that there is indeed a tightly bound \insitu~ ``iceberg'', whose tips have been recognized before. The fact that there are many metal-poor stars in the inner galaxy has also already been recognized by the recent work of \citet{Arentsen2020a,Arentsen2020b}. 
%DW This double cite did not format right. VC: fixed. 
% However, those authors did not carry out spatial mapping, orbit analysis, or an analysis of abundance patterns.

Our analysis here is in many ways preliminary and suggests various avenues of follow-up. The spatial distribution of this population deserves to be modeled quantitatively, including the effects of dust extinction, source crowding, parallax uncertainties, and the giant luminosity function. The abundances, and perhaps more importantly, the abundance patterns, deserve full spectroscopic follow-up to understand: How pure are the XP selected samples in this regime? Is there a floor in the \MH\ distribution that might reflect the \MH\ of the proto-Milky Way's circumgalactic medium? If these are indeed among the most ancient stars in the Milky Way, do they have remarkable abundance ratios? Was the formation of this component associated with the formation of an extensive set of globular clusters? What can the degree of central concentration of this component tell us about the violence of subsequent merging events, which would scatter stars to larger radii?

On a more technical side, this analysis reflects the astounding information content of the Gaia DR3 data, particularly the XP spectra. Our data-driven approach to estimate \MH\ seems to work well for the current analysis, at the price of restricting \MH\ estimates to bright objects. Presumably, we are far from exploiting the information of the XP spectra, which should be unlocked by forward-modeling of the data.

\newpage
%\begin{acknowledgments}

%We thank our colleague Daniela Ruz-Mieres for her quick and invaluable help with using GaiaXPy.
% We are grateful to
% Rohan Naidu
% for valuable conversations and feedback. 
VC gratefully acknowledges a Peirce Fellowship from Harvard University. 
The authors thank Vasily Belokurov and Andrey Kravtsov for helpful comments on the manuscript.

This work has made use of data from the European Space Agency (ESA) mission
{\it Gaia} (\url{https://www.cosmos.esa.int/gaia}), processed by the {\it Gaia}
Data Processing and Analysis Consortium (DPAC,
\url{https://www.cosmos.esa.int/web/gaia/dpac/consortium}). Funding for the DPAC
has been provided by national institutions, in particular the institutions
participating in the {\it Gaia} Multilateral Agreement.

Funding for the Sloan Digital Sky 
Survey IV has been provided by the 
Alfred P. Sloan Foundation, the U.S. 
Department of Energy Office of 
Science, and the Participating 
Institutions. 

SDSS-IV acknowledges support and 
resources from the Center for High 
Performance Computing  at the 
University of Utah. The SDSS 
website is www.sdss.org.

SDSS-IV is managed by the 
Astrophysical Research Consortium 
for the Participating Institutions 
of the SDSS Collaboration including 
the Brazilian Participation Group, 
the Carnegie Institution for Science, 
Carnegie Mellon University, Center for 
Astrophysics | Harvard \& 
Smithsonian, the Chilean Participation 
Group, the French Participation Group, 
Instituto de Astrof\'isica de 
Canarias, The Johns Hopkins 
University, Kavli Institute for the 
Physics and Mathematics of the 
Universe (IPMU) / University of 
Tokyo, the Korean Participation Group, 
Lawrence Berkeley National Laboratory, 
Leibniz Institut f\"ur Astrophysik 
Potsdam (AIP),  Max-Planck-Institut 
f\"ur Astronomie (MPIA Heidelberg), 
Max-Planck-Institut f\"ur 
Astrophysik (MPA Garching), 
Max-Planck-Institut f\"ur 
Extraterrestrische Physik (MPE), 
National Astronomical Observatories of 
China, New Mexico State University, 
New York University, University of 
Notre Dame, Observat\'ario 
Nacional / MCTI, The Ohio State 
University, Pennsylvania State 
University, Shanghai 
Astronomical Observatory, United 
Kingdom Participation Group, 
Universidad Nacional Aut\'onoma 
de M\'exico, University of Arizona, 
University of Colorado Boulder, 
University of Oxford, University of 
Portsmouth, University of Utah, 
University of Virginia, University 
of Washington, University of 
Wisconsin, Vanderbilt University, 
and Yale University.

This work made use of the Third Data Release of the GALAH Survey (Buder et al. 2021). The GALAH Survey is based on data acquired through the Australian Astronomical Observatory, under programs: A/2013B/13 (The GALAH pilot survey); A/2014A/25, A/2015A/19, A2017A/18 (The GALAH survey phase 1); A2018A/18 (Open clusters with HERMES); A2019A/1 (Hierarchical star formation in Ori OB1); A2019A/15 (The GALAH survey phase 2); A/2015B/19, A/2016A/22, A/2016B/10, A/2017B/16, A/2018B/15 (The HERMES-TESS program); and A/2015A/3, A/2015B/1, A/2015B/19, A/2016A/22, A/2016B/12, A/2017A/14 (The HERMES K2-follow-up program). We acknowledge the traditional owners of the land on which the AAT stands, the Gamilaraay people, and pay our respects to elders past and present. This paper includes data that has been provided by AAO Data Central (datacentral.org.au).

The Guoshoujing Telescope (the Large Sky Area Multi-Object Fiber Spectroscopic Telescope LAMOST) is a National Major Scientific Project built by the Chinese Academy of Sciences. Funding for the project has been provided by the National Development and Reform Commission. LAMOST is operated and managed by the National Astronomical Observatories, Chinese Academy of Sciences.

This work has made use of the Python package GaiaXPy, developed and maintained by members of the Gaia Data Processing and Analysis Consortium (DPAC), and in particular, Coordination Unit 5 (CU5), and the Data Processing Centre located at the Institute of Astronomy, Cambridge, UK (DPCI).

%\end{acknowledgments}

\appendix 
\section{ \MH ~Estimates via XGBoost trained on SDSS APOGEE Abundances}

Here we expand on a few specifics of the \MH ~estimate that is summarized in Section~\ref{sec:MH-dist}.

\subsection{Training of XGBoost model of [M/H]}

We want to derive \MH from XP information for objects that still have significant flux in the blue ($G_{\text{BP}}<15.5$), yet may be reddened by several magnitudes. To help break any degeneracies between extinction and T$_{\text{eff}}$ we include near-infrared photometry from ALLWISE \citep{2014yCat.2328....0C}. We adopt SDSS DR17 APOGEE abundances \citep{2022ApJS..259...35A} as a training sample, since it covers the inner disk and contains many giants stars with high extinctions.

For the machine learning algorithm, we choose the extreme gradient boosting algorithm \citep[][hereafter XGBoost]{Chen:2016:XST:2939672.2939785}, as it is computationally inexpensive to train and can outperform other algorithms, e.g. Deep Learning \citep[e.g.][]{https://doi.org/10.48550/arxiv.2207.08815}. To improve the \MH\ predictions for our sample (consisting of red giant stars by construction), we also restrict the training sample from SDSS DR17 sample to only giants, via $\log g<3.5$.

This leaves us with a suitable choice of data features, for which we explored a number of options. Our starting point was the set of XP coefficients \citep{DeAngeli2022,2021A&A...652A..86C}, which we normalized by the Gaia apparent $G$ flux. In Table~\ref{table:xgboost-test-error-coefficients}, we quote the test errors for using the different numbers of XP coefficients. The performance is overall very good but depend only weakly on how many coefficients we use. This suggests that also the high-order coefficients contain useful information, and, at least for this application and the APOGEE DR17 sample, a truncation of XP coefficients is neither required nor helpful in our case.

\begin{table}
\caption{Test errors of [M/H] from XGBoost from 20-fold cross-validation using various numbers of XP coefficients. We quote the median absolute error and the root-mean-square error.}
\label{table:xgboost-test-error-coefficients}
\centering
\begin{tabular}{lcc}
input features & median AE & RMSE \\
\hline\hline
all 55 coefficients &     0.061  &   0.145 \\
first 40 coefficients only &        0.059  &   0.138 \\
first 30 coefficients only &       0.060   &  0.137 \\
first 20 coefficients only &        0.062  &   0.139 \\
first 10 coefficients only &       0.066  &   0.144 \\
\hline
XP colours only &       0.078   &  0.154   \\
all coefficients + XP colours  &  0.064  &   0.138 \\
\hline
all coefficients + WISE &      0.088       &  0.186 \\
XP colours + WISE &           0.067   &  0.136 \\
all coefficients + XP colours + WISE &  0.049   &  0.107 \\
\hline
% all coefficients + XP colours + WISE &  0.052  &   0.118 \\
\end{tabular}
\end{table}

Given the results in \citet{Montegriffo2022b}, we then attempted to only use photometry synthesized from XP spectra using GaiaXPy\footnote{\url{https://gaia-dpci.github.io/GaiaXPy-website/}} and combine this with ALLWISE near-infrared photometry. Specifically, we synthesize photometry from the systems of Pristine, StromgrenStd, Jpas, Jplus, SkyMapper, and HstAcswfc. As we can see from Table~\ref{table:xgboost-test-error-coefficients}, using only such colors has a similar performance as only using the XP coefficients, despite including ALLWISE photometry. However, a substantial improvement is obtained when combining all XP coefficients with these colors.

In the end, we used GaiaXPy to compute the following synthetic photometry and combined it with all 110 {\tt XP} coefficients in the XGBoost training on SDSS DR17 (APOGEE):

\begin{align*}
G - W_1 \\
G_\textrm{BP}-G_\textrm{RP} \\
G_\textrm{BP}-W_2 \\
W_1-W_2 \\
\textrm{CaHK}_\textrm{Pristine} - W_1 \\
\textrm{CaHK}_\textrm{Pristine} - v_\textrm{StromgrenStd} \\
\textrm{CaHK}_\textrm{Pristine} - b_\textrm{StromgrenStd} \\
\textrm{CaHK}_\textrm{Pristine} - y_\textrm{StromgrenStd} \\
\textrm{Jplus}_g - \textrm{Jplus}_i \\
\textrm{Jplus}_{0395} - \textrm{CaHK}_\textrm{Pristine} \\
\textrm{Jplus}_{0515} - \textrm{CaHK}_\textrm{Pristine} \\
\textrm{Jplus}_{0861} - \textrm{CaHK}_\textrm{Pristine} \\
G_\textrm{RP} - \textrm{Jplus}_i \\
v_\textrm{StromgrenStd} - 2\cdot b_\textrm{StromgrenStd} + y_\textrm{StromgrenStd} \\
\textrm{CaHK}_\textrm{Pristine} - 2\cdot \textrm{Jplus}_{0410} + \textrm{Jplus}_{0430} \\
\end{align*}
We emphasize that the $G_\textrm{BP}-W_2$ color has the longest wavelength leverage and therefore is highly affected by extinction, whereas the $W_1-W_2$ color is comparatively insensitive to extinction.

\begin{figure*}
\begin{center}
\includegraphics[width=5.7cm]{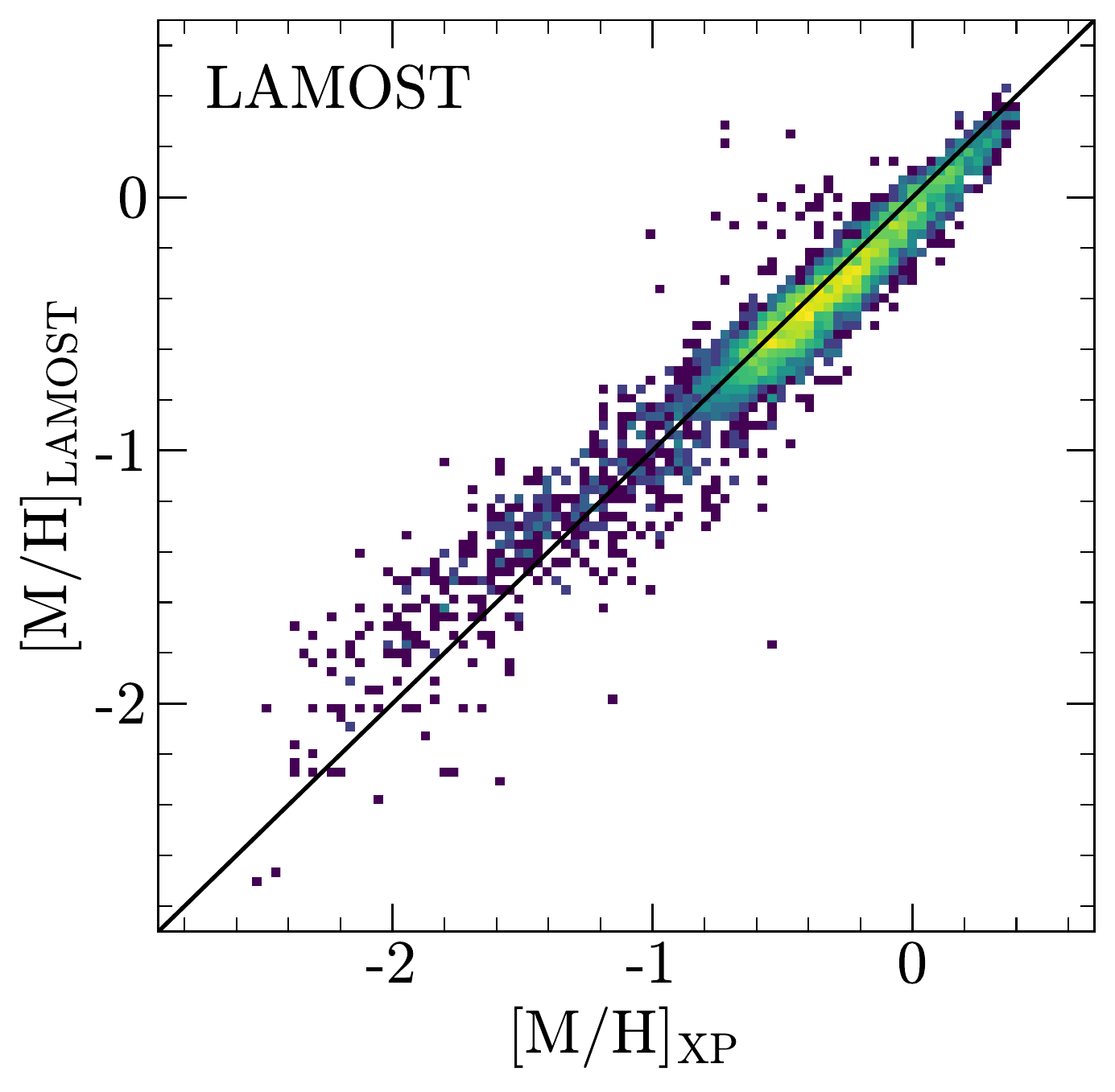}
\includegraphics[width=5.7cm]{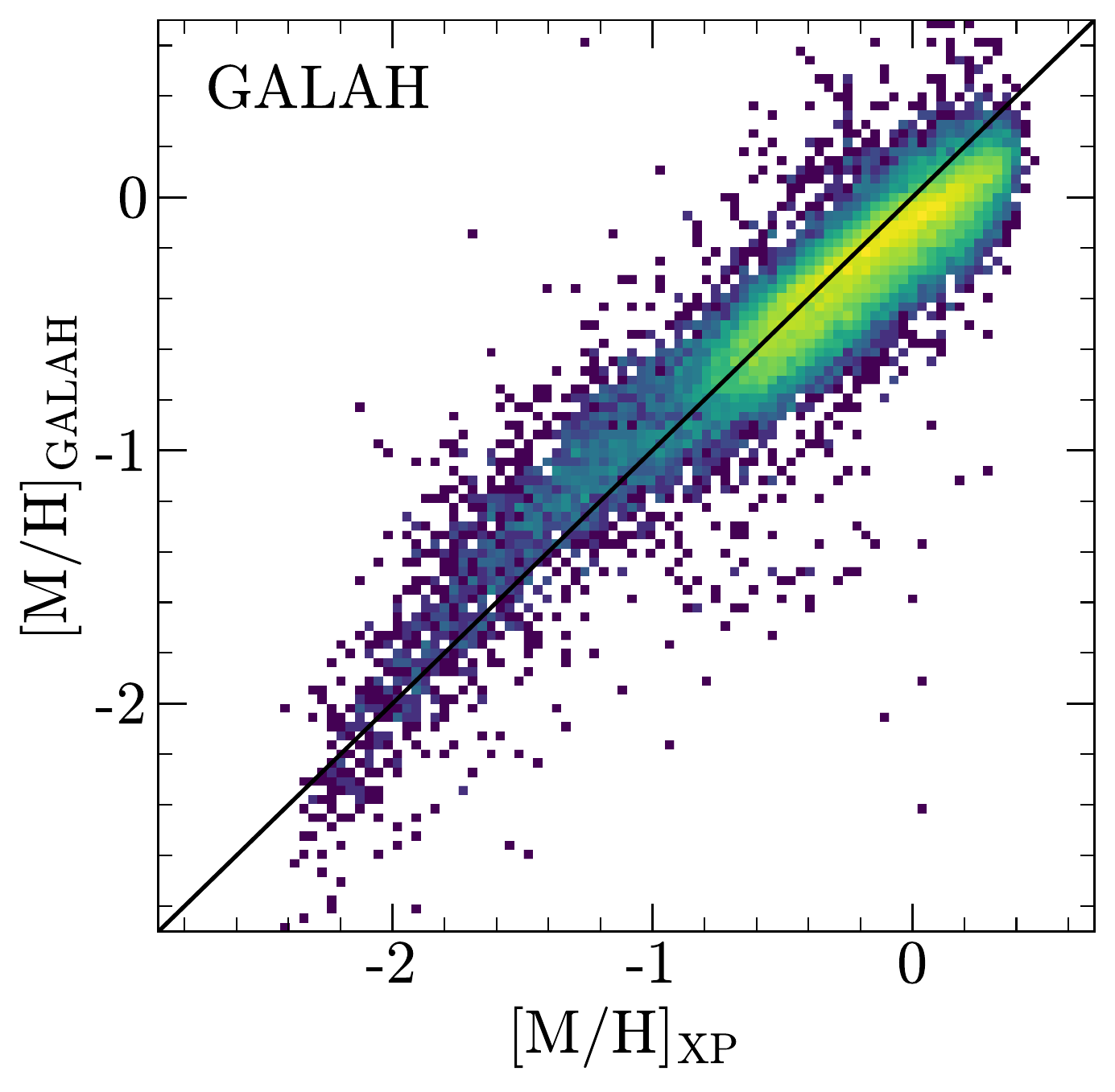}
\includegraphics[width=5.7cm]{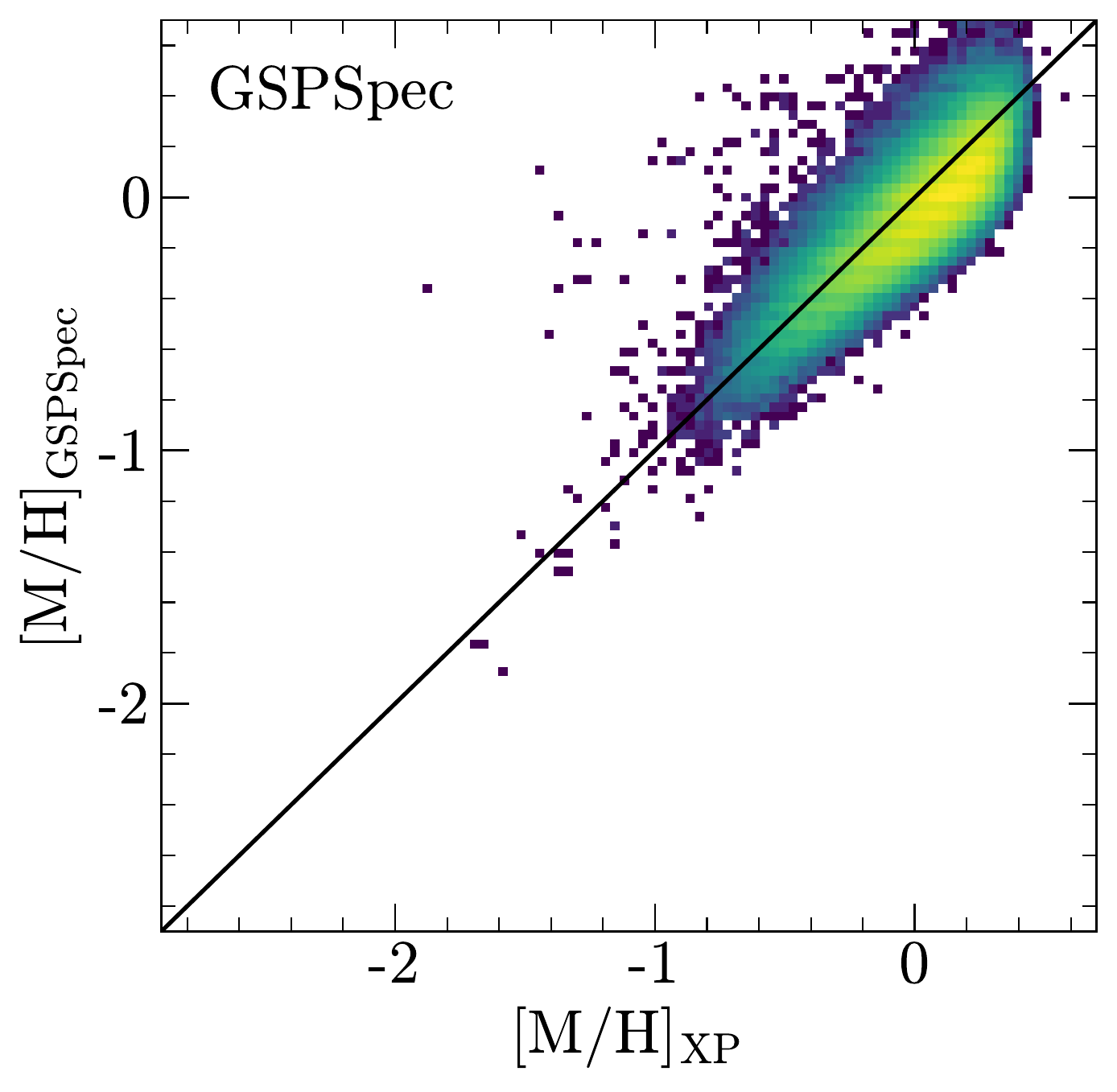}
\end{center}
\caption{Cross-validation of our \MH$_{\text{XP}}$~ estimates against three external spectroscopic data sets, LAMOST \citep[here][]{Xiang2019} , GALAH \citep{GALAH_DR3} and {\tt GSP-Spec} \citep{Recio-Blanco-RVS-abundances}. This cross-validation supports the precision and robustness of our \MH$_{\text{XP}}$ estimates.}
\label{fig:MH_cross_validation}
\end{figure*}

% \begin{figure}

% \end{figure}

As is evident from Table~\ref{table:xgboost-test-error-coefficients}, this final configuration has slightly worse performance than the best configuration. However, the median absolute errors are similar, that is, most sources get similar results. The only noteworthy difference is in the RMS error, suggesting that the final configuration may have a few more outliers. Still, the results are better than what we can achieve from coefficients alone. 

\subsection{Further [M/H] Validation}\label{sec:more_mhvalidation}

\begin{figure}
\includegraphics[width=\columnwidth]{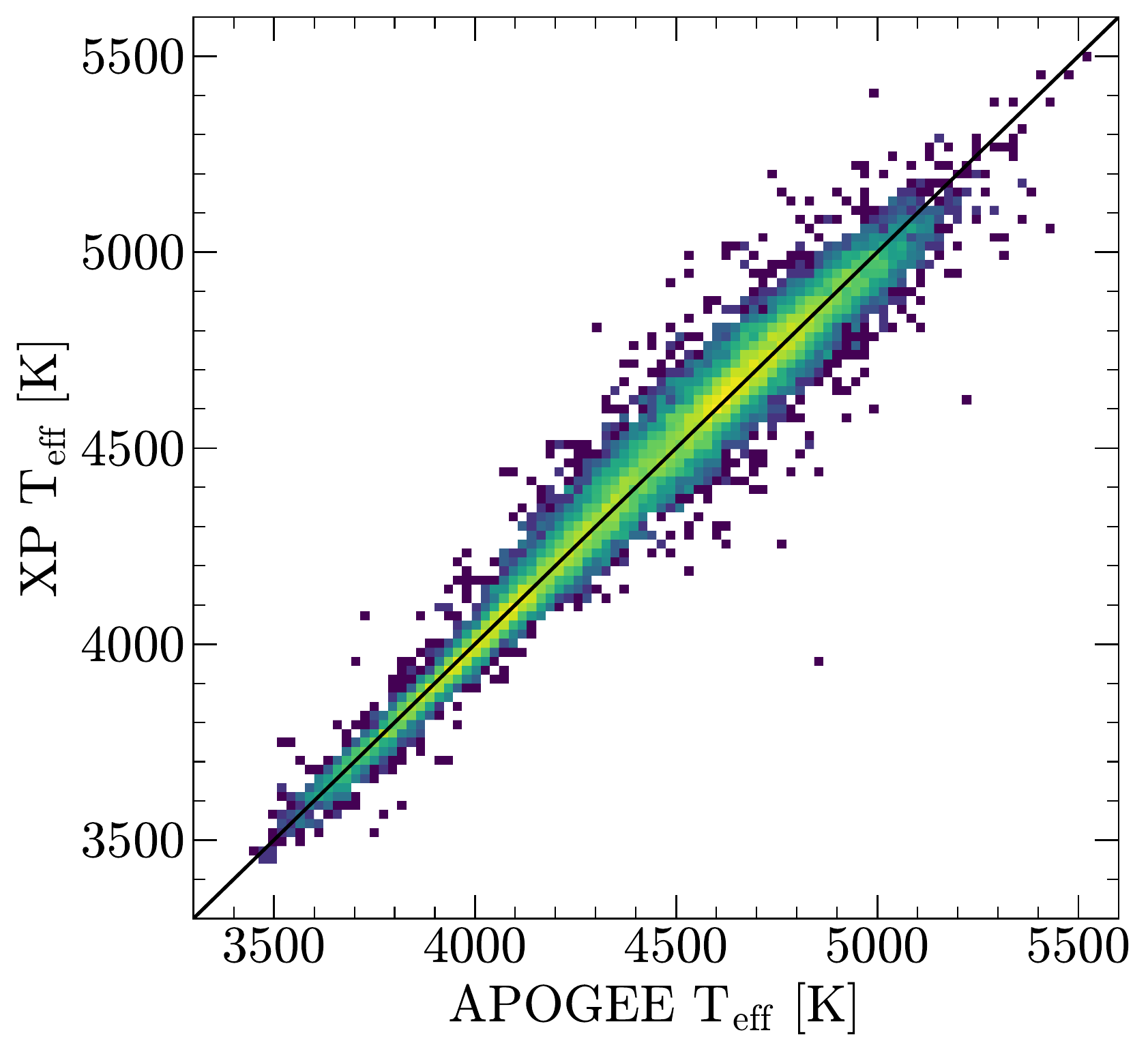}
\includegraphics[width=\columnwidth]{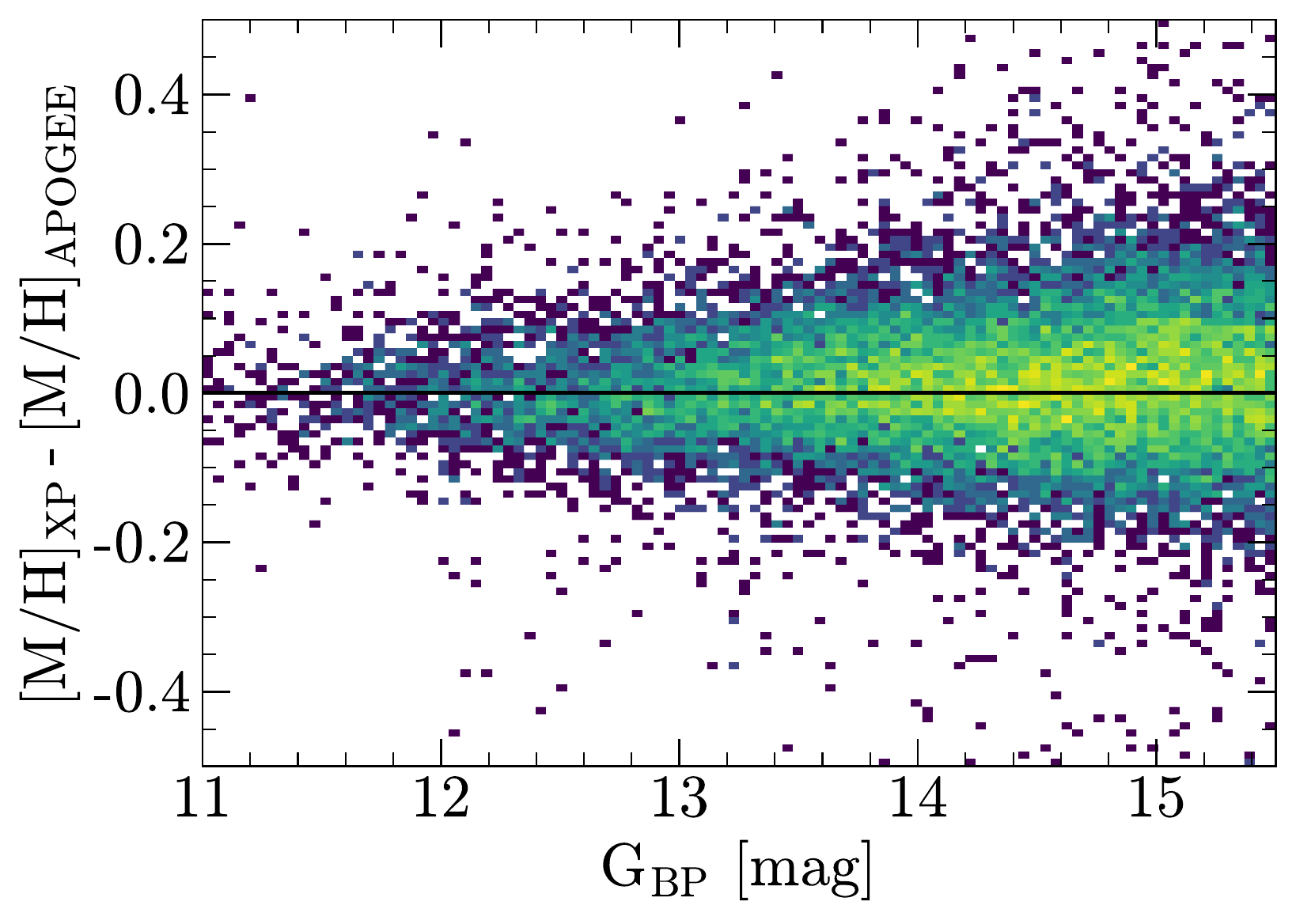}
\caption{Top: Validation of our T$_{\text{eff}}$(XP) estimates against the T$_{\text{eff}}$ from SDSS DR17 APOGEE. The XP predictions match the APOGEE values within a median $\Delta$T$_{eff}=32$~K, and with a mean difference of only 3~K. Bottom: Validation of our \MH ~estimates as a function of G$_{\text{BP}}$ magnitude, with $\Delta\MH \equiv \MH_{\text{XP}} - \MH_{\text{APOGEE}}$, showing that the quality of the estimate depends somewhat on G$_{\text{BP}}$, as expected, but is overall unbiased.}
\label{fig:BP_validation}
\label{fig:Teff_validation}
\end{figure}

Figure~\ref{fig:MH_cross_validation} illustrates a cross-validation comparison of our results to other spectroscopic surveys. Such cross-validation is easy to interpret if one can assume that the external data set represents ``ground truth''. We applied a {\tt SNR\_G}~$>25$ to the LAMOST sample and required that there were no quality (concern) flags in any of the GALAH spectra.  The left panel shows that the \MH$_{\text{XP}}$ estimates are accurate for LAMOST, which itself is tied to the APOGEE \MH\ scale, with a slight increase in scatter towards low \MH. The comparison with GALAH (middle panel) affirms the precision and purity of the low-\MH$_{\text{XP}}$ estimates. Note that GALAH has a different scaling between \MH\ and [Fe/H] than APOGEE and LAMOST, which may explain the offset of the estimates for low-\MH ~stars from the 1-to-1 line. In the parent sample without stringent GALAH quality cuts, there are many stars for which GALAH implies that they are $\MH<-1$, but the \MH$_{\text{XP}}$ estimate disagrees. The position of those stars in the de-reddened color-magnitude plane suggests that they indeed have $\MH >-1$. In other words, for these cases \MH$_{\text{XP}}$ appears to be the more \emph{robust} estimate.  The comparison with GSP-Spec in the right panel of Figure~\ref{fig:MH_cross_validation} confirms the quality of GSP-Spec's \MH\ estimates. But it also shows that the GSP-Spec values do not cover low \MH\  \citep[by design, see][]{Recio-Blanco-RVS-abundances}.

%Clearly, the XGBoost performance is excellent at the metal-poor end. Quantitatively, for $\textrm{[M/H]}<-1$ the completeness is 67.5\% and contamination is 8.5\%. While the completeness will allow us to efficiently identify metal-poor stars in our application sample, the contamination rate estimated from our APOGEE sample cannot be generalized to the application sample, and we will have to revisit this specifically. 

Because a star's effective temperature (T$_{\text{eff}}$) can be very helpful in sample selection and characterization, we also estimate T$_{\text{eff}}$ using the same XP features and XGBoost trained on SDSS DR17 APOGEE T$_{\text{eff}}$, as illustrated in Fig.~\ref{fig:Teff_validation}. The XGBoost predictions based on XP spectra and WISE photometry match the APOGEE values within a median $\Delta$T$_{\text{eff}}=32$~K, and with a mean difference of only 3~K. Figure~\ref{fig:BP_validation} compares our [M/H] estimates to APOGEE as a function of Gaia $G_{\text{BP}}$ magnitude, showing the slight degradation towards the faintest sources in the sample.

We also compared our \MH~ estimates to the extensive set of photometric estimates for metal-poor stars from \citet{2021ApJS..254...31C} based on SkyMapper DR2 (SM2) photometry. We find a mean \MH~ offset of 0.28~dex, and for stars with $\MH_{\text{SM2}}\sim -1.5$ a central 68\% interval of $p\bigl (\,  \MH_{\text{SM2}} - \MH_{\text{XP}}\, \bigr )$ of 1~dex. Clearly, the cross-validation of our XP-based \MH~ estimates and SkyMapper DR2 shows far more scatter than our cross-validation with external spectroscopic data sets. This illustrates that for this sample XP-based \MH~ estimates outperform those of SM2.

% \section{Conversion between [$\alpha$/M] and [$\alpha$/Fe]}

% Some spectroscopic surveys provide estimates of [M/H] and [$\alpha$/M] \citep[e.g.][]{} whereas others provide estimates of [Fe/H] and [$\alpha$/Fe] \citep[e.g.][]{}. In this appendix, we briefly recap how to convert between those estimates. For the conversion from [Fe/H] to \MH\, we use the following relation \citep[e.g.][Sect.~8.1 therein]{2005essp.book.....S}
% %
% \begin{equation}
% \MH \sim \textrm{[Fe/H]} + \log_{10}\left(0.694\cdot 10^{[\alpha/Fe]} + 0.306\right)
% \end{equation}
% %
% This is equivalent to:
% %
% \begin{equation}
% \textrm{[M/Fe]} = \MH - \textrm{[Fe/H]} \sim \log_{10}\left(0.694\cdot 10^{[\alpha/Fe]} + 0.306\right)
% \end{equation}
% %
% We thus obtain:
% %
% \begin{equation}
% [\alpha/Fe] - [\alpha/M] = \textrm{[M/Fe]} \sim \log_{10}\left(0.694\cdot 10^{[\alpha/Fe]} + 0.306\right)
% \end{equation}
% %
% such that
% %
% \begin{equation}
% [\alpha/M] \sim [\alpha/Fe] -  \log_{10}\left(0.694\cdot 10^{[\alpha/Fe]} + 0.306\right)
% \end{equation}

\clearpage

\bibliography{bib}
\bibliographystyle{aasjournal}

\end{document}